\documentclass[manuscript,screen]{acmart}
\usepackage{blindtext}
\AtBeginDocument{%
  \providecommand\BibTeX{{%
    \normalfont B\kern-0.5em{\scshape i\kern-0.25em b}\kern-0.8em\TeX}}}
\usepackage{multirow}
\usepackage{amsmath}
\usepackage{relsize}
\usepackage{graphicx} 
\usepackage{colortbl}
\usepackage{amsmath}
\usepackage{caption,subcaption}
\usepackage{tablefootnote}
\setcopyright{acmcopyright}
\copyrightyear{2022}
\acmYear{2022}
\acmDOI{XXXXXXX.XXXXXXX}

\definecolor{silver}{rgb}{0.753,0.753,0.753}
\definecolor{mgray}{rgb}{0.128,0.128,0.128}
\begin{document}
\title[Retrieval for Extremely Long Queries and Documents with RPRS]{Retrieval for Extremely Long Queries and Documents with RPRS: a Highly Efficient and Effective Transformer-based Re-Ranker}%
\author{Arian Askari}
\orcid{0000-0003-4712-832X}
\email{a.askari@liacs.leidenuniv.nl}
\affiliation{%
  \institution{Leiden University}
  \country{The Netherlands}
}
\author{Suzan Verberne}
\orcid{0000-0002-9609-9505}
\email{s.verberne@liacs.leidenuniv.nl}
\affiliation{%
  \institution{Leiden University}
  \country{The Netherlands}
}
\author{Amin Abolghasemi}
\orcid{0009-0002-3725-7312}
\email{m.a.abolghasemi@liacs.leidenuniv.nl}
\affiliation{%
  \institution{Leiden University}
  \country{The Netherlands}
}
\author{Wessel Kraaij}
\orcid{0000-0001-7797-619X}
\email{w.kraaij@liacs.leidenuniv.nl}
\affiliation{%
  \institution{Leiden University}
  \country{The Netherlands}
}
\author{Gabriella Pasi}
\orcid{0000-0002-6080-8170}
\email{gabriella.pasi@unimib.it}
\affiliation{%
  \institution{University of
Milano-Bicocca}
  \country{Italy}
}

\renewcommand{\shortauthors}{Askari and Verberne, et al.}

\begin{abstract}
Retrieval with extremely long queries and documents is a well-known and challenging task in information retrieval and is commonly known as Query-by-Document (QBD) retrieval. Specifically designed Transformer models that can handle long input sequences have not shown high effectiveness in QBD tasks in previous work. We propose a \textbf{R}e-Ranker based on the novel \textbf{P}roportional \textbf{R}elevance \textbf{S}core (RPRS) to compute the relevance score between a query and the top-k candidate documents. Our extensive evaluation shows RPRS obtains significantly better results than the state-of-the-art models on five different datasets. Furthermore, RPRS is highly efficient since all documents can be pre-processed, embedded, and indexed before query time which gives our re-ranker the advantage of having a complexity of $O(N)$ where $N$ is the total number of sentences in the query and candidate documents. Furthermore, our method solves the problem of the low-resource training in QBD retrieval tasks as it does not need large amounts of training data, and has only three parameters with a limited range that can be optimized with a grid search even if a small amount of labeled data is available. Our detailed analysis shows that RPRS benefits from covering the full length of candidate documents and queries.
\end{abstract}


\maketitle
\section{Introduction}
Query-by-document (QBD) retrieval is a task in which a seed document acts as a query -- instead of a few keywords -- with the aim of finding similar (relevant) documents from a document collection \cite{yang2009query,yang2018retrieval,le2021query}. Examples of QBD tasks are professional, domain-specific retrieval tasks such as legal case retrieval \cite{yoshioka2021coliee,tran2020encoded,askari2021combining,shao2020bert}, patent prior art retrieval \cite{piroi2011clef,piroi2019multilingual,fujii2007overview}, and scientific literature retrieval \cite{cohan2020specter,mysore2021csfcube}. In the literature, three other terms are used to refer to this type of task: Query-by-Example (QBE) retrieval, document similarity ranking, and document-to-document similarity ranking.
\par
Transformer-based ranking models \cite{vaswani2017attention}, such as BERT-based rankers \cite{nogueira2019passage,yan2019idst} have yielded improvements in many IR tasks. However, the time and memory complexity of the self-attention mechanism in these architectures is $O(L^2)$ for a sequence of length $L$  \cite{kitaev2020reformer}. For that reason, architectures based on BERT have an input length limitation of 512 tokens. 
This causes challenges in QBD tasks where we have long queries and documents. For instance, the average length of queries and documents in the legal case retrieval task of COLIEE 2021 \cite{askari2021combining} is more than $5k$ words. Variants of Transformers that aim to cover long sequences such as LongFormer \cite{beltagy2020longformer} and Big Bird \cite{zaheer2020big} have not shown high effectiveness, which could be due to either their scattered attention mechanism \cite{sekulic2020longformer} or to the limited number of training instances for QBD tasks \cite{askari2021combining}. 
To overcome these limitations, in this paper we propose an effective and efficient Transformer-based re-ranker for QBD retrieval that covers the whole length of the query and candidate documents without memory limitations. We focus on re-ranking in a two-stage retrieval pipeline following prior studies that address retrieval tasks in a multi-stage retrieval pipeline \cite{nogueira2019passage,padigela2019investigating,AbolghasemiECIR22,chen2022incorporating}. 
\par
In domain-specific tasks, there is often the need for expensive professional searchers as annotators in order to create QBD test collections, e.g., lawyer annotators for case law retrieval tasks. This makes the annotation process for QBD tasks expensive and, as a result, low-resource training sets are a significant issue in these tasks \cite{askari2021combining}. One such example is the training set of the COLIEE 2021 dataset for case law retrieval that consists of only $3,297$ relevant documents for $900$ queries as training instances, which is very few compared to general web search datasets, e.g. MS MARCO \cite{nguyen2016ms} has approximately $161$ times more training instances ($532,761$ queries). This limitation is overcome by our proposed method, RPRS (a \textbf{R}e-ranker based on \textbf{P}roportional \textbf{R}elevance \textbf{S}core computation), which 
does not need large amounts of training data because it has three parameters with a limited range that can be -- but do not necessarily need to be -- optimized using a grid search if a small amount of labelled data is available. Compared to BM25, our proposed method has only one more parameter, which we optimize over 10 different possible values. The two other parameters' ranges are equal to those of BM25.
\par
Given a query document and a set of candidate documents, we split the query and document text into sentences and use Sentence-BERT (SBERT) bi-encoders \cite{reimers2019sentence} to embed the individual sentences.
Then, a relevance score is computed by the RPRS relevance model we propose in this paper, which is based on the similarity of individual sentences between a query and a document.
The intuition behind the proposed method is based on the following assumption: a candidate document from the ranked list of documents in response to a query is more likely to be relevant if it contains a relatively high number of sentences that are similar to query sentences compared to other documents'
sentences in the ranked list. %
Since RPRS has been designed based on this assumption, we conduct an experimental investigation to assess the quality of our proposed method by addressing the following research questions:
\par
\textbf{RQ1:} What is the effectiveness of RPRS compared to the State-of-the-art models for QBD retrieval?
\par
\noindent To answer this question we evaluate the effectiveness of RPRS on legal case retrieval  (i.e. given a legal case, find the related cases) using the COLIEE 2021 dataset \cite{goebel2021coliee}. As RPRS uses a SBERT model in order to embed query and document sentences, we first run a series of experiments to find the SBERT model with which RPRS achieves the highest effectiveness. Additionally, we investigate two different unsupervised and self-supervised domain adaption approaches on the top-two most effective SBERT models without domain adaption. We conclude that RPRS achieves higher effectiveness with domain-specific Transformers in the SBERT architecture 
when adapting them to the domain by using TSDA \cite{wang2021tsdae}. We continue the experiments based on this finding.
We also compare RPRS to two prior models: (1) the state-of-the-art model on the COLIEE dataset: MTFT-BERT \cite{AbolghasemiECIR22}, and (2) Self-Supervised Document Similarity Ranking (SDR) \cite{ginzburg-etal-2021-self} which also uses bi-encoders for sentence-level representation to compute relevance scores and is the state-of-the-art on two Wikipedia document similarity datasets.
\par 
For relevance estimation with RPRS, we split both the query and document content into sentences. However, there are other ways for splitting query and document. Therefore, we investigate if sentences are the most appropriate and informative textual units for RPRS by addressing:
\par
\textbf{RQ2:} How effective is RPRS with shorter or longer text units instead of sentences?
\par
\noindent Next, for analyzing the cross-data generalizability of the proposed method we address the following question:
\par
\textbf{RQ3}: What is the effectiveness of RPRS with parameters that were tuned on a different dataset in the same domain?
\par
\noindent To address this question, we have evaluated the proposed method on the \emph{Caselaw} dataset \cite{locke2017automatic} with the parameters that were tuned on the COLIEE dataset without doing domain adaption on the collection for the Transformer model. In other words, our goal is to analyze how much the three optimized parameters of RPRS on the COLIEE dataset are transferable to another dataset in the same (i.e., legal) domain.  
Next, we assess the generalizability of our method by addressing:
\par 
\textbf{RQ4:} To what extent is RPRS effective and generalizable across different domains with different type of documents?
\par
\noindent In this regard, we evaluate our method on two different domains using datasets of patents and Wikipedia webpages. For patents, we use CLEF-IP 2011~\citep{piroi2011clef}, a patent prior art retrieval dataset, and for Wikipedia we assess the effectiveness of our method on WWG and WWA datasets, which are two datasets for similarity ranking between Wikipedia pages in the wine and video game domains. We show that the proposed method achieves higher ranking effectiveness over the state-of-the-art in all five datasets (COLIEE, Caselaw, CLEF-IP 2011, WWG, WWA).\footnote{The implementation is available on \url{https://github.com/arian-askari/rprs}.} %
\par
In summary, our contributions are as follows:
\begin{enumerate}
    \item We propose an effective and highly efficient re-ranker (RPRS) for QBD tasks that covers the full text of query and candidate documents without any length limitation. The efficiency of our method is justified by using bi-encoders and its effectiveness is evaluated on five datasets. 
    \item RPRS is suitable for QBD retrieval datasets with low-resource training data since it has only three parameters with limited range that can be optimized using a grid search even if a relatively low amount of labelled data is available. %
    \item We show how the use of various SBERT models and adapting them on domain specific datasets affects the effectiveness of RPRS. This results in three SBERT models in the legal, patent, and Wikipedia domains, which we will make publicly available on Huggingface.%
    \item Our proposed model outperforms the state-of-the-art models for each of the five benchmarks, including SDR \cite{ginzburg-etal-2021-self}, which was also proposed as a sentence-based ranker for QBD tasks.
    \item An ablation study shows that Document Proportion -- the proportion of the document that is relevant to the query -- is the most important component in RPRS. In addition, we found the crucial role of covering the full length of queries and documents for QBD tasks by studying the effect of feeding RPRS with truncated queries and documents.
\end{enumerate}
\par
The structure of the paper is as follows: after a discussion of related work in Section \ref{sec:related}, we describe the proposed method in section \ref{sec:proposed_method} and the details of the BM25 and SDR baselines in section \ref{sec:baselines}. The experiments and implementation details are covered in Section \ref{sec:experiments}. The results are examined and the research questions are addressed in Section \ref{sec:results}. In section \ref{sec:discussion}, we investigate and discuss our proposed method in more-depth. Finally, the conclusion is described in Section \ref{sec:conclusion}.
\section{Related Work}\label{sec:related}
In the following, we first introduce Query-by Document (QBD) retrieval tasks. Then, we provide an overview on prior methods for sentence embedding and pre-training. Finally, we give an overview of previous works on sentence-level retrieval methods.
\subsection{QBD Tasks}
\textbf{Legal case retrieval.} 
In countries with \emph{common law} systems, finding supporting precedents to a new case, is vital for a lawyer to fulfill their responsibilities to the court. However, with the large amount of digital legal records -- the number of filings in the U.S. district courts for total cases and criminal defendants was and is 544,460 in 2020\footnote{\url{https://www.uscourts.gov/statistics-reports/judicial-business-2020}} -- it takes a significant amount of time for legal professionals to scan for specific cases and retrieve the relevant sections manually. Studies have shown that attorneys spend approximately 15 hours in a week seeking case law \cite{lastres2015rebooting}. This workload necessitates the need for information retrieval (IR) systems specifically designed for the legal domain. The goal of these systems is to assist lawyers in their duties by exploiting AI and traditional information retrieval methods. One particular legal IR task is case law retrieval.
\par
The majority of the work on case law retrieval takes place in COLIEE, the Competition on Legal Information Extraction and Entailment \cite{rabelo2020coliee,goebel2021coliee}. \citet{askari2021combining} combine lexical and neural ranking models for legal case retrieval. They optimize BM25 and obtain state-of-the-art results with lexical models. However, recently, on top of the optimized BM25, \citet{AbolghasemiECIR22} present multi-task learning as a re-ranker for QBD retrieval and set the new state-of-the-art. The limitation of the method by \cite{AbolghasemiECIR22} is that the input is limited by the BERT architecture (512 tokens) \cite{devlin2018bert}. Therefore, it cannot cover the full-text of long documents in QBD datasets like COLIEE.
\par
\noindent \textbf{Patent prior art retrieval.}
Patents serve as stand-ins for various domains including technological, economic, and even social activities. The Intellectual Property (IP) system encourages the disclosure of innovative technology and ideas by granting inventors exclusive monopoly rights over the economic value of their inventions.
Therefore, according to Piroi et al., \cite{piroi2011clef} patents have a considerable impact on the market value of companies. With the number of patent applications filed each year continuing to rise, the need for effective and efficient solutions for handling such huge amounts of information grows more essential.
There are various patent analysis tasks. In this work, we focus on patent prior art retrieval, the aim of which is to find patent documents in the target collection that may invalidate a specific patent application \cite{piroi2011clef,piroi2019multilingual}.  
\par
Patent retrieval is a challenging task: the documents are lengthy, the language is formal, legal, and technical with long sentences~\cite{verberne2010quantifying}. Shalaby et al. \cite{shalaby2019patent} provide a detailed review of research on patent retrieval that shows that the most successful methods on the CLEF-IP \cite{piroi2011clef} benchmarks are traditional lexical-based methods~\cite{piroi2013overview,verberne2009prior}, sometimes extended with syntactic information~\citep{d2011combining}. There are more recent studies that focus on patent retrieval in passage-level assessment \cite{hofstatter2019enriching,althammer2021cross}. However, recent studies do not address document-level patent retrieval %
due to the limitation of Transformer-based methods on taking into account the full length of the documents. \citet{mahdabi2014query} propose an effective approach for patent prior art retrieval, which is based on collecting a citation network using a specific API that is not provided by the organizers of the CLEF-IP dataset. Taking into account the fact that this setup is different from the original setup, we do not consider this approach as our baseline for a fair comparison.
\par
\noindent \textbf{Document similarity for Wikipedia-based datasets.} Estimating the similarity within Wikipedia pages is useful for many applications such as clustering, categorization, finding relevant web pages, etc. \citet{ginzburg-etal-2021-self} propose two new datasets on Wikipedia annotated by experts. The documents come from the wine and video game domains; we refer to the collections as (1) Wikipedia video games (WVG) and (2) Wikipedia wine articles (WWA). For both datasets, the task is finding relevant Wikipedia pages given a seed page.

\subsection{Sentence embedding}
Sentence embedding is a well-studied topic and it is suitable for measuring sentence similarity, clustering, information retrieval via semantic search, etc.
BERT-based sentence embedding methods employ Transformer models to effectively and efficiently embed sentences. The challenge is that BERT's architecture makes it inappropriate for semantic similarity search in its original configuration since it requires both sentences to be concatenated into the network and applies multiple attention layers between all tokens of both sentences, which has a significant computational overhead. For instance, to find the most similar pair in a collection of 10,000 sentences BERT needs roughly 50 million inference calculations, which takes around 65 hours \cite{reimers2019sentence}.
\par
\citet{humeau2019poly} introduce poly-encoders to tackle the run-time overhead of the BERT cross-encoder and present a method to calculate a score employing attention between the context vectors and pre-computed candidate embeddings. However, Poly-encoders have the disadvantage that their score function is not symmetric and that their computing cost is excessively high and require $O(n^2)$ score calculations.
\citet{reimers2019sentence} therefore proposed Sentence-BERT (SBERT) as a variant of the pre-trained BERT model that uses siamese and triplet network architectures to produce semantically meaningful embeddings that can be compared using cosine similarity. SBERT decreases the time it takes to find the most similar pair from $65$ hours to around $5$ seconds while keeping BERT's performance \cite{reimers2019sentence}. In this work, we utilize SBERT as our embedding model, making the efficiency of RPRS much higher than cross-encoder based re-rankers. We exploit a selection of SBERT models that are tuned on different datasets.
\par
It is noteworthy to mention that while our proposed methods leverage bi-encoders, specifically Sentence-BERT \citet{reimers2019sentence}, to re-rank the first-stage retriever's ranked list, our approach fundamentally diverges from dense passage retrievers that employ bi-encoders  as first stage retrievers in terms of both methodology and application\cite{karpukhin-etal-2020-dense}. Dense retrievers emphasize on the alignment of query and relevant document representations by optimizing a targeted loss function, such as the mean square error between the query and relevant document vectors while we do not have such training in our proposed. Moreover, optimizing dense retrievers on the query-by-document task with extremely lengthy queries or documents is not computationally possible due to BERT model's maximum word limit of 512 words which prevents from representing each query or document by a single pass to a bi-encoder, and as a result, prevents from being able to optimizing the representation of query and document as they do not have one union representation. There could be future work on studying how applying dense passage retrieval on query by document task which is out of the scope of this study.
\subsection{Unsupervised sentence embeddings training}
Previous works on unsupervised sentence embeddings learning have achieved promising results on semantic textual similarity tasks by combining pre-trained Transformers with various training objectives. \citet{carlsson2020semantic} propose Contrastive Tension (CT) that views identical and different sentences as positive and negative examples respectively and trains two independent encoders. BERT-flow \cite{li2020sentence} debiases the embedding distribution towards Gaussian to train the model. SimCSE \cite{gao2021simcse} uses contrastive learning \cite{hadsell2006dimensionality,chen2020simple} to classify identical sentences with different dropout masks as positive instances.
\par
\citet{wang2021tsdae} propose an unsupervised state-of-the-art method called TSDAE that is based on Transformers and denoising auto-encoders that 
encode damaged sentences into fixed-sized vectors and require the decoder to reconstruct the original sentences from these sentence embeddings \cite{wang2021tsdae}.
\citet{wang2021gpl} propose Generative Pseudo Labeling (GPL), which combines a query generator with pseudo labeling from a cross-encoder. However, the GPL methodology is not suitable for QBD tasks as it relies on cross-encoders that are limited both in length and efficiency. We evaluate the effect of two state-of-the-art pre-training methods in section \ref{sec:results}.

\subsection{Sentence-level Retrieval}
Addressing relevance of candidate documents by leveraging sentence-level evidence has been studied for long documents retrieval in the past years \cite{yilmaz2019applying,akkalyoncu-yilmaz-etal-2019-cross,yilmaz2019cross,hofstatter2021intra,li2023power}. However,
there is no work on sentence-based relevance score computation using Transformer-based models (i.e,  SBERT \cite{reimers2019sentence}) on QBD retrieval tasks similar to our approach. \citet{yilmaz2019applying} apply inference at the sentence level for each document and aggregate the sentence-level inference by learning a weight for each top-scoring sentence in each candidate document. \citet{zhang2017uwaterloomds} observe that the `best' sentence or paragraph in a document gives a decent proxy for document relevance, which was the inspiration for \citet{yilmaz2019applying} as well. We also use this intuition in our approach to QBD Retrieval, where both the query and the documents are long texts. \citet{mysore2021multi} propose a scientific document similarity model based on sentence-level similarity that leverages co-citation sentences as a source of document similarity.
\par
Recently, \citet{ginzburg-etal-2021-self} proposed an unsupervised ranker called Self-Supervised Document Similarity Ranking (SDR) that computes the final relevance score by computing two sentence- and paragraph-level matrices. They evaluate SDR's effectiveness on two new datasets annotated by human experts. We replicate SDR \cite{ginzburg-etal-2021-self} as the most recent and comparable methodology to RPRS because: (1) Although SDR's mechanism is dissimilar to RPRS fundamentally, it also is a sentence-level relevance scoring model designed for QBD tasks that covers the full length of both queries and candidate document texts using sentence embeddings, and (2) Similar to RPRS, SDR has a complexity of $O(N)$ where $N$ is count of sentences -- due to the utilization of bi-encoder sentence embeddings. Therefore, besides comparing RPRS with the state-of-the-art model on each dataset, we compare its effectiveness to SDR. SDR's architecture makes it suitable for both full-ranking and re-ranking settings.
\begin{figure*}[t]
  \scalebox{.70}{\includegraphics{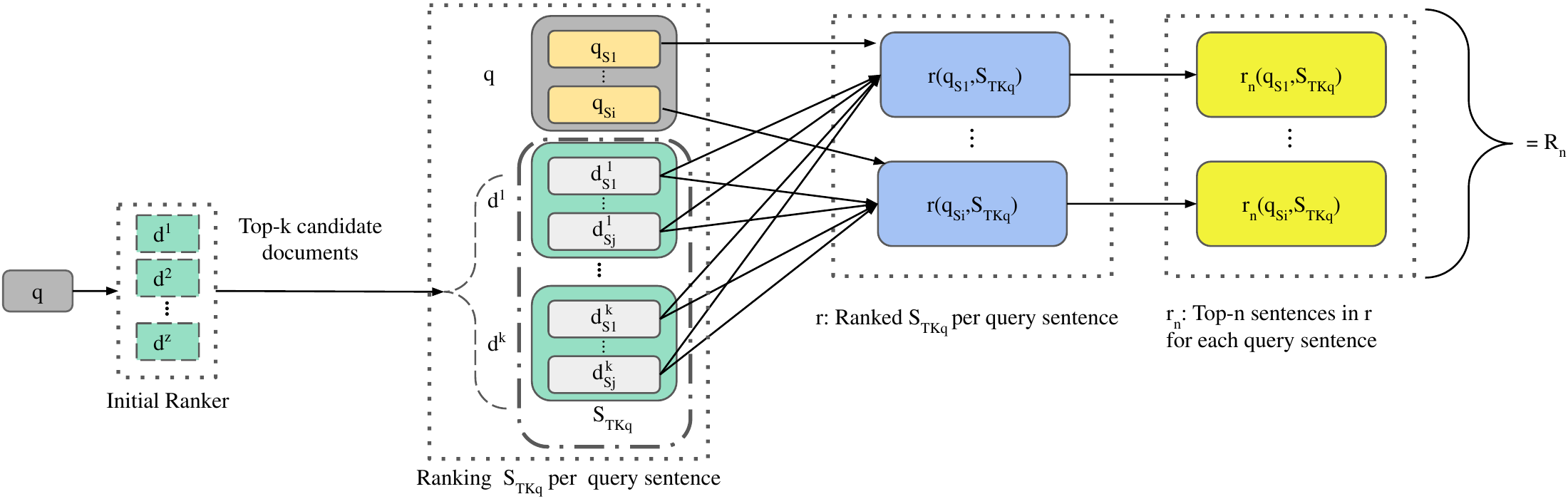}}%
  \caption{The workflow of computing $R_n$ given a query document $q$ and set of top-$k$ candidate documents. $i$ and $j$ refer to the last sentence of each query and document.}%
  \label{fig:rn}
\end{figure*}
\section{Proposed Method: RPRS}\label{sec:proposed_method}
As mentioned, we assume that a candidate document $d$ is likely to be relevant to a query document $q$ if a large proportion of $d$ is similar to $q$, and a large proportion of $q$ is similar to $d$. 
A specific challenge in QBD tasks is that a document can be very long (more than $10$k words) and it may contain several topics. For a long query document $q$, this can cause an irrelevant candidate document $d$ positioned on top because only one topic of $d$ is very similar to one topic of $q$. Similarly, a long irrelevant candidate document could be ranked on top becaus one of its topics is very similar to only one topic of query document. We address this problem in our method by integrating the length of the query and document into equations \ref{eq:qp} and \ref{eq:dp} that we elabore in the following.
In the following, we define our concepts and proposed methodology.
\subsection{Definitions}
\label{sec:notation}
In Table \ref{tab:symbold} we show the legend of the symbols we employ in order to formally introduce our methodology. 
In our sentence-based relevance model, we first retrieve the set of documents $TK_q$ for $q$ with an initial ranker. Given $q$, we compute the cosine similarity of each $\overrightarrow{q_{s}}$ with each  $\overrightarrow{d_{s}}$ , i.e. we compute the similarity of each query sentence with each sentence in $S_{TK_q}$.
The result $r(q_{s},S_{TK_q})$ is a ranked list of sentences from $S_{TK_q}$ for $q_{s}$.
The set $r_n(q_{s},S_{TK_q})$ contains the top-$n$ sentences from $r(q_{s},S_{TK_q})$, i.e. the document sentences most similar to $q_{s}$. From now on, we call a $d_s$ `\textit{most similar}' to $q_s$ if $d_s$ is a member of $r_n(q_{s},S_{TK_q})$, meaning that it is among the top-n sentences from the ranked list for sentences from the top-k candidate documents. %
This indicates that the document sentence $d_s$ is placed among the top-$n$ highest most similar document sentences to the query sentence $q_s$.

The set $R_n(S_{q},S_{TK_q})$ contains all $r_n(q_{s},S_{TK_q})$ for all query sentences.
Figure \ref{fig:rn} shows the workflow of computing $R_n$ step-by-step. Based on our assumption, $d$ and $q$ are likely to be relevant if a large proportion of $S_{d}$ occurs in a large proportion of each $r_n$s from $R_n$.
\par
\begin{table}[]
\caption{Meaning of symbols}
\begin{tabular}{l|l}
\hline
\rule[-2ex]{0pt}{0pt}\rule{0pt}{3ex}\textbf{Symbol} & \textbf{Meaning} \\ \hline

\rule[-2ex]{0pt}{0pt}\rule{0pt}{3ex}q and d & query and candidate document \\ \hline
\rule[-2ex]{0pt}{0pt}\rule{0pt}{3ex}$q_{s}$ and $d_{s}$ & a sentence of $q$ and $d$ respectively\\ \hline
\rule[-2ex]{0pt}{0pt}\rule{0pt}{3ex}$\overrightarrow{q_{s}}$ and  $\overrightarrow{d_{s}}$ & vector representation (sentence embedding) of $q_{s}$ and $d_{s}$\\ \hline
\rule[-2ex]{0pt}{0pt}\rule{0pt}{3ex}$S_d$ and $S_q$ & the sets of all sentences of $d$ and $q$. \\ \hline
\rule[-2ex]{0pt}{0pt}\rule{0pt}{3ex}$TK_q$ & set of top-$k$ candidate documents retrieved for $q$ \\ \hline
\rule[-2ex]{0pt}{0pt}\rule{0pt}{3ex}$S_{TK_q}$ & set of all sentences from $TK_q$ \\ \hline
\rule[-2ex]{0pt}{0pt}\rule{0pt}{3ex}$r(q_{s},S_{TK_q})$ & a ranked list of $S_{TK_q}$ for $q_{s}$ \\ \hline
\rule[-2ex]{0pt}{0pt}\rule{0pt}{3ex}$r_n(q_{s},S_{TK_q})$ & top-$n$ sentences from $r(q_{s},S_{TK_q})$ \\ \hline
\rule[-2ex]{0pt}{0pt}\rule{0pt}{3ex}$R_n(S_{q},S_{TK_q})$ & contains all $r_n(q_{s},S_{TK_q})$ for all sentences of $S_q$\\ \hline

\end{tabular}
\label{tab:symbold}
\end{table}
\subsection{Re-ranker based on Proportional Relevance Score (RPRS)}\label{sec:proposed_method_pb} 
We formally define the proportional  relevance score for $RPRS$ as:
\begin{equation}
\label{eq:prs}
    \small
    \begin{array}{c}
        RPRS(q,d,S_{TK_q},n) 
        = QP(S_q,S_d,S_{TK_q}, n) \times DP(S_d,S_q,S_{TK_q}, n)
    \end{array}
\end{equation}
Where $PRS$ is the Proportional Relevance Score (PRS), $QP$ is Query Proportion, and $DP$ is Document Proportion.
Given $q$, we compute $PRS$ for $d$ based on the $R_n$ and $S_{TK_q}$.
The parameter $n$ controls the number of top-$n$ similar sentences per query sentence in $r_n$ and $R_n$.
In the following, we define two functions: $q_{R_n}(S_q,S_d,S_{TK_q})$ and $d_{R_n}(S_d,S_q,S_{TK_q})$ that we use in the later equations for computing $QP$ and $DP$:
\begin{equation}
\label{eq:qrn}
    q_{R_n}(S_q,S_d,S_{TK_q}) = 
    \sum_{q_{s}}^{S_q} min(1, |S_d \cap r_n(q_{s},S_{TK_q})|)
\end{equation}
Where $|x|$ denotes the cardinality
of set $x$, and the $min$ function returns 1 if at least one of the sentences of $d$ ($S_d$'s sentences) is in $r_n(q_{s},S_{TK_q})$, zero otherwise.
Here, we do not take into account the repetition of $S_d$'s sentences in each $r_{n}$ for a $q_{s}$ because of the $min$ function. However, we incorporate that in a controllable way in the extended variation ($RPRS w/freq$) in section \ref{sec:freq} by defining parameter $k1$.
The function $q_{R_n}$ counts for how many sentences of $q$, at least one sentence of the candidate document sentences ($S_d$) occurs at least one time in the set of $q_{s}$'s $r_n$. Next, we define $d_{R_n}$ which is the main component for computing $DP$ as:
\begin{equation}
\label{eq:drn}
    d_{R_n}(S_d,S_q,S_{TK_q}) = \mathlarger{\mathlarger{\sum_{d_{s}}^{S_d}}} min(1, \sum_{r_n}^{R_n} | \{d_{s}\} \cap r_n|) 
\end{equation}
\par
Where $\{d_{s}\}$ denotes a singleton, i.e. a set with only one sentence of $S_d$. 
For a query document $q$ and a candidate document $d$, the function $d_{R_n}$ iterates over all sentences of $S_d$ and counts how many sentences of $d$ occur at least one time in $R_n$.
Given the parameter $n$, it is possible that more than one sentence of $d$ occurs in $r_n$ for a query sentence.
We argue that equation \ref{eq:qrn} and \ref{eq:drn} are complementary to each other as each equation assess the relevance from either the query or candidate document perspective. We show the impact of each equation by the ablation study in section \ref{sec:ablation}.
We now define $QP$ and $DP$ based on equations (\ref{eq:qrn}) and (\ref{eq:drn}):
\par
\begin{equation}
\label{eq:qp}
    QP(S_q,S_d,S_{TK_q}, n) = \frac{q_{R_n}(S_q,S_d,S_{TK_q})}{\text{count of $q$'s sentences}}%
\end{equation}
\begin{equation}
\label{eq:dp}
    DP(S_d,S_q,S_{TK_q}, n) =  \frac{d_{R_n}(S_d,S_q,S_{TK_q})}{\text{count of $d$'s sentences}}
\end{equation}
It is noteworthy that the denominator of $QP$ (count of $q'$s sentences) has the same value for all candidate documents, and thus it could be ignored for ranking; however, we keep it as it makes relevance scores comparable for score analysis.
\par
Intuitively, the relevance score of $d$ should increase by having more of its sentences in $R_n$ for query $q$. 
The advantage of this design is that the candidate document $d$ receives the highest relevance score if all $r_n$s' sentences in $R_n$ are from $d$. On the other hand, $d$ receives the lowest relevance score if none of its sentences are in $r_n$s' sentences in $R_n$.
However, there is a disadvantage in this design as it does not consider the repeated appearances of candidate document sentences in $r_n$'s. Therefore, we propose another variation on RPRS called RPRS w/freq that takes into account this in a controllable way explained in the next section.
\par
\subsection{Taking into account frequency (RPRS w/freq)}\label{sec:freq}
In the previous section, by using the minimum function in equations \ref{eq:qrn} and \ref{eq:drn}, we only accounted for the occurrence 
of \emph{at least one} of the candidate documents' sentences in  $r_n$ and $R_n$.
In other words, we did not consider the frequency of the occurrences in our definition for functions $q_{R_n}$ and $d_{R_n}$. To empower our model, we define $Fq_{R_n}$ and $Fd_{R_n}$ as the modified versions of $q_{R_n}$ and $d_{R_n}$ that take into account the frequency: 
\begin{equation}\label{eq:fqrn}
    Fq_{R_n}(S_q,S_d,S_{TK_q}) = \sum_{q_{s}}^{S_q} \frac{|S_d \cap r_n(q_{s},S_{TK_q})|}{|S_d \cap r_n(q_{s},S_{TK_q})|+k1((1-b)+\frac{b \cdot dl}{avgdl})}
\end{equation}
\begin{equation}\label{eq:fdrn}
    Fd_{R_n}(S_d,S_q,S_{TK_q}) = \sum_{d_{s}}^{S_d} 
    \frac{\sum_{r_n}^{R_n} | \{d_{s}\} \cap r_n|}{\sum_{r_n}^{R_n} | \{d_{s}\} \cap r_n|+k1((1-b)+\frac{b \cdot dl}{avgdl})}
\end{equation}
Here, inspired by the BM25 Okapi schema, we take the frequency into account in a controllable way.
To this aim, we introduce a modified version of RPRS, which we denote by RPRS w/freq, which relies on two parameters: $k1$, the frequency saturation parameter \footnote{To gain a deeper insight into k1 and frequency saturation, we recommend referring to the original BM25 paper \citet{robertson1994some}. For a more accessible explanation of frequency saturation and its effects, you can explore the details provided in \cite{understandingbm25}. This resource offers a step-by-step breakdown that should make it easier to understand how this parameter works.}, for controlling the effect of frequency of occurrence of a candidate document's sentence in $R_n$, and $b$, the document length normalization parameter, for controlling the effect of the candidate document length. Similar to the BM25 formula, $dl$ refers to document length and $avgdl$ refers to average length of documents. 
The advantage of the frequency saturation parameter ($k1$) of the proposed method which works similarly to BM25's `term saturation' mechanism, is that the occurrence of multiple sentences of a candidate document in only one $r_n$ of $R_n$ has a lower impact on the relevance score than the occurrence of multiple sentences of a candidate document each of which occurs only once in several different $r_n$s of $R_n$. 
This characteristic is in line with BM25's concept of frequency saturation, which prevents a document from obtaining a high score solely based on the repetition of a single word that matches just one word from the query.
We parameterize the degree of normalizing the relevance according to the document length (i.e., number of sentences of a document) with the parameter $b$, which works similarly to BM25's length normalization parameter \cite{robertson1994some,lipani2015verboseness}. If $b=0$, the relevance score is not normalized by the document length at all because the right side of the denominator in equation \ref{eq:fqrn} and \ref{eq:fdrn} will be: $k1((1-b)+\frac{b \cdot dl}{avgdl}) = ((1-0)+\frac{0 \cdot dl}{avgdl}) = k1((1)+0) = k1$. Therefore, only $k1$ will be kept in the denominator and the whole denominator will be $|S_d \cap r_n(q_{s},S_{TK_q})|+k1$ and $\sum_{r_n}^{R_n} | \{d_{s}\} \cap r_n|+k1$ for equation \ref{eq:fqrn} and \ref{eq:fdrn} respectively. As $b$ increases from $0$ towards $1$, the impact of document length -- compared to the average length of documents: $dl/adl$ -- in normalizing the relevance score will be higher. Consequently, $b=1$ means the relevance score will be fully normalized based on the document length because the right side of the denominator in equation \ref{eq:fqrn} and \ref{eq:fdrn} will be: $k1((1-b)+\frac{b \cdot dl}{avgdl}) = ((1-1)+\frac{1 \cdot dl}{avgdl}) = k1((0)+\frac{1 \cdot dl}{avgdl}) = k1 \cdot \frac{dl}{avgdl}$. Therefore, the document length will be divided by the average length of documents and will be multiplied to $k1$. As a result, the denominator will be fully normalized based on the ratio of document length to the average length of all documents in the denominator, and the whole denominator will be $|S_d \cap r_n(q_{s},S_{TK_q})|+k1 \cdot  \frac{dl}{avgdl}$ and $\sum_{r_n}^{R_n} | \{d_{s}\} \cap r_n|+k1 \cdot  \frac{dl}{avgdl}$ for equations \ref{eq:fqrn} and \ref{eq:fdrn} respectively.
\par
In summary, the final proposed method has three parameters: $k1$ and $b$ as described above, and $n$ that controls the number of top-$n$ sentences in $r_n$ per query sentence. All parameters can be tuned on the training set. Furthermore, the parameters can be used with default values or with values that are obtained by tuning the method on another dataset. With \emph{RPRS} and \emph{RPRS w/freq}, we only need to compute the cosine similarity between embeddings of query sentences and the document sentences. Our proposed method is efficient because all documents can be pre-processed, embedded and indexed before query time. At query time, producing the embedding for query sentences using SentenceBERT is highly efficient \cite{reimers2019sentence}. This gives either \emph{RPRS} or \emph{RPRS w/freq} the advantage of having a complexity of $O(N)$ where $N$ is the total number of sentences in the query and candidate documents compared to re-rankers based on Cross-encoders with $O(N^2)$.
Furthermore, calculating cosine similarity is a simple operation, thus resulting in a fast inference time.

\section{Baseline retrieval models}\label{sec:baselines}
In this section we introduce BM25 and SDR that are the two main baselines in our experiments.
\subsection{BM25}
Lexical retrievers estimate the relevance of a document to a query based on word overlap \cite{robertson2009probabilistic}.
Many lexical methods, including vector space models, Okapi BM25, and query likelihood, have been developed in previous decades. We use BM25 because of its popularity as first-stage ranker in current systems and its strong effectiveness on QBD tasks \cite{rosa2021yes}. Based on the statistics of the words that overlap between the query and the document, BM25 calculates a score for the pair:
\begin{equation}
s_{lex}(q,d) = BM25(q,d) = \sum_{t \in q \cap d }{rsj_t . \frac{tf_{t,d}}{tf_{t,d} + k_{1} \{ (1-b) + b \frac{|d|}{l} \} }}
\end{equation}
where $t$ is a term, $tf_{t,d}$ is the frequency of $t$ in document $d$, $rsj_t$ is the Robertson-Spärck Jones weight \cite{robertson1994some} of $t$, and $l$ is the average document length. $k_1$ and $b$ are parameters.
\subsection{Birch}
\citet{yilmaz2019applying} present a simple yet effective solution for applying BERT to long document retrieval: the inference is applied to each sentence in a candidate document, and sentence-level evidence is aggregated for ranking documents as follows:
\begin{equation}
Score_d = a \cdot S_{doc} + (1-a) \cdot \sum_{i=1}^{n} w_i\cdot S_i
\end{equation}
Where $S_{doc}$ represents the original document score, and $S_i$ denotes the $i$-th top-scoring sentence according to BERT. The parameters $a$ and $w_i$'s can be learned or used with default values. To replicate the `3S: BERT' models, referred to as 3S-Birch hereafter, we utilize the official implementation from the Birch paper, which uses the three top-scoring sentences \footnote{\url{https://github.com/castorini/birch}}. We report the 3S-Birch approach, as it yielded the highest effectiveness, even though we experimented with the top-1 and top-2 scoring sentences. Until now, the effectiveness of Birch has only been analyzed and proven for short queries and long documents. In contrast, our investigation focuses on evaluating its performance in situations where both queries and documents are extremely long. We use the same BERT model that we use for our proposed method per each dataset.
\subsection{SDR \cite{ginzburg-etal-2021-self}}
We replicate SDR \cite{ginzburg-etal-2021-self} as the most recent and comparable methodology to RPRS because: (1) Although SDR's mechanism is dissimilar to RPRS fundamentally, it also is a sentence-level relevance scoring model designed for QBD tasks that cover the full length of both queries and candidate document texts using sentence embeddings, and (2) Similar to RPRS, SDR has a complexity of $O(N)$ where $N$ is count of sentences -- due to the utilization of bi-encoder sentence embeddings instead of cross-encoders -- for computing relevance scores in contrast to other techniques with $O(N^2)$ \cite{devlin2018bert,barkan2020scalable}.
Therefore, besides comparing RPRS with the state-of-the-art model on each dataset, we compare its effectiveness with SDR. SDR's architecture makes it suitable for both full-ranking and re-ranking settings; as a result, we evaluate it in both configurations and only report the result of the setup (ranker or re-ranker) in which it performs best. In the following, we introduce SDR's pre-training and inference methodology.
\subsubsection{Self-supervised pre-training}
Given a collection of documents $D$, SDR samples sentence pairs from the same paragraph of a given document (intra-samples) and sentence pairs from different paragraphs taken from different documents (inter samples) with equal probability (i.e., $0.5$ for each type of sampling). %
It tokenizes sentences and aggregates them into batches, and it randomly masks them in a similar way to the RoBERTa pre-training paradigm. The authors use the Roberta model for their implementation \cite{liu2019roberta}.
The pretraining objective of SDR comprises a dual-term loss: (1) a standard MLM loss adopted from \cite{devlin2018bert} that allows the model to specialize in the domain of the given collection of documents \cite{gururangan2020don},
and (2) a contrastive loss \cite{hadsell2006dimensionality} aims to minimize the distance between the representations of sentence-pairs from the same paragraph (intra-samples), while maximizing the distance between the representations of sentence-pairs from different paragraphs (inter-samples).
\subsubsection{Inference}\label{sec:sdr_ranker}
SDR computes the relevance score between a query $q$ and a candidate document $d$ by computing a two-stage hierarchical similarity score: 
(1) SDR creates a sentence similarity matrix $M$ for all possible pairs of query paragraphs and candidate document paragraphs. Each cell in $M$ represents the cosine similarity between a sentence from paragraph $i$ of the query and a sentence from paragraph $j$ of the candidate document. Specifically, the rows of $M$ correspond to sentences from the query paragraph, and the columns correspond to sentences from the candidate document paragraph. Therefore, a cell $M_{a,b}^{i,j}$ represents the cosine similarity between sentence $a$  from paragraph $i$ of the query and sentence $b$ from paragraph $j$ of the candidate document; (2) Next, a paragraph similarity matrix $P$ is created for the candidate document $d$ based on all pairs of paragraphs (query paragraph vs document paragraph). Each cell in $P$ represents the similarity between a paragraph from the query and a paragraph from the candidate document and is computed using the maximum cosine similarity between sentences in the two paragraphs one from the query and one from the candidate document.
The motivation for the creation of matrix $P$ is that similar paragraph pairs should incorporate similar sentences that are more likely to correlate under the cosine metric.
\par
Finally, $P$ is normalized globally ($NRM(P)$), based on other candidate documents' paragraph similarity matrices.
Based on the $NRM(P)$ generated for a candidate document, %
the total similarity score $S$ is obtained by computing an average of the highest cosine similarity scores in each row of $P$.
The motivation for $S$ is that the most correlated paragraph pair only contributes to the total similarity score. In contrast to RPRS, SDR does not take into account the length of candidate document because the denominator in its formula for computing matrix $P$ is count of sentences in a query's paragraph and for total score $S$ is the  count of paragraphs in the query. 
However, this issue has been handled in our re-ranker by integrating the length of the query and document into $QP$ and $DP$ (See section \ref{sec:proposed_method_pb}). Moreover, if one paragraph of a candidate document be the most similar paragraph to all paragraph of query, SDR does not penalize this repetition while we take that into account with $k1$ parameter in \emph{RPRS w/freq} that controls frequency saturation (See section \ref{sec:freq}). In addition, 
the cosine score is used directly into the approach, while we only consider that score for ranking sentences in $R_n$ (See section \ref{sec:proposed_method_pb} and Figure \ref{fig:rn}).
\section{Experiments}\label{sec:experiments}
In this section, we first describe the datasets, the pre-trained Transformers that we use, and our implementation details for the retrieval models. Finally, we provide information about parameter tuning and pre-processing of the dataset.
\subsection{Datasets}
We evaluate our models on three QBD retrieval tasks: legal case retrieval, patent prior art retrieval, and document similarity ranking for Wikipedia pages.
\paragraph{\textbf{COLIEE'21.}}
We first use the COLIEE'21 dataset to evaluate the effectiveness of the proposed method (PRS) on legal case retrieval.
There are $650$ query documents in the train set and $250$ in the test set \cite{goebel2021coliee}, with 4,415 documents as candidate documents in both sets. 
The candidate documents' average length is $5,226$ (tokenized by SparkNLP \cite{KOCAMAN2021100058}, see below), with outliers reaching $80,322$ words.
\paragraph{\textbf{Caselaw.}}
To investigate the generalizability of our model, \emph{RPRS w/freq}, we tuned it on the COLIEE'21 dataset and evaluate the tuned model on a different dataset: the Caselaw dataset \cite{locke2017automatic}. Therefore, we use the Caselaw dataset as a test set to assess how well the tuned \emph{RPRS w/freq} model can generalize to new and unseen data.
The Caselaw dataset contains $100$ query documents, $2,645$ relevance assessments, and $63,916$ candidate documents. 
In Caselaw, the candidate documents' average length is $2,344$, with outliers up to $124,092$ words.
\paragraph{\textbf{CLEF-IP 2011.}}
For patent prior art retrieval \cite{piroi2011clef}, we experiment on the CLEF-IP 2011 dataset that contains $300$ and $3,973$ query documents in the train set and test set respectively, both of which have document-level relevance assessments. The count of candidate documents per query is about $3$ Million. In this work, we select the English subset of CLEF-IP 2011 that contains about $900,000$ candidate documents per query, and $100$ and $1,324$ query documents in the train set and test set respectively. 
We concatenate title, abstract, description, and claims as the whole patent document.
The candidate documents' average length in CLEF-IP is $10,001$ words, with outliers up to $407,308$ words.
\paragraph{\textbf{Wikipedia.}}
For the Wikipedia datasets, we use the Wikipedia video games (WVG) and the Wikipedia wine articles (WWA) datasets \cite{ginzburg-etal-2021-self} that contain $21,935$ and $1,635$ candidate documents and $90$ and $92$ query documents respectively. The WWG collection consists of articles reviewing video games from all genres and consoles and the WWA collection consists of a mixture of articles discussing different types of wine categories, brands, wineries, grape varieties, and more. %
The documents' average length is $1061$ and $966$, with outliers up to $23,048$ and $13,081$ words for the WVG and WWA datasets respectively.
\subsection{\textbf{Baselines}}
BM25 has previously been shown to be a strong baseline for QBD retrieval \cite{rosa2021yes}, and it holds the state-of-the-art among all lexical models on COLIEE \cite{askari2021combining}.
We implement BM25 as the initial ranker using Elasticsearch on all five datasets.
In addition to BM25 on the complete query document text, we employ BM25 on the top-$10$ percent of query document terms that are extracted and scored 
using Kullback-Leibler divergence for Informativeness (KLI)~\citep{verberne2016evaluation} following prior work \cite{locke2017automatic,askari2021combining}. As KLI has been shown to be an effective approach for making shorter queries for BM25 \cite{locke2017automatic}, we employ KLI on documents of all three datasets and refer to that in the result tables as `BM25 + KLI'.

Moreover, we replicate the SDR \cite{ginzburg-etal-2021-self} ranker as $SDR_{inf}$ where `inf' refers to inference. $SDR_{inf}$ is a comparable sentence-level baseline to \emph{RPRS w/freq}.
It should be noted that wherever SDR is mentioned in the tables, it is referring to the best-performing SDR variant, either as a re-ranker after the best first-stage ranker, or as a full ranker.
We experiment with Roberta besides the other SBERT's models as SDR's authors use the Roberta model for their implementation \cite{liu2019roberta}.
\par
For COLIEE 2021, we compare the proposed method to the lexical state-of-the-art model, which is $BM25_{optimised}$ with optimized parameters ($b=1, k=2.8$) \cite{askari2021combining}. Additionally, we do comparisons with two re-rankers: the BERT re-ranker \cite{nogueira2019passage}, and the multi-task fine-tuned BERT re-ranker (MTFT-BERT), which is the neural state-of-the-art on COLIEE \cite{AbolghasemiECIR22}.
For Caselaw \cite{locke2017automatic}, we re-use the lexical and neural models that we fine-tuned on COLIEE to analyze the generalizability of them and the proposed method.
The state-of-the-art method for Caselaw is \emph{K} \cite{locke2017automatic}.\footnote{We do not report result with manually created boolean queries for fair comparison.} We found that \emph{K}'s run file is the most effective initial ranker and thus we used that as our initial ranker on Caselaw \cite{locke2017automatic}.
\par %
For patent retrieval, there is no Transformer-based method baseline, which could be due to the fact that average length of documents in the patent dataset is around $10,001$ and thus cannot be handled by Transformer models straightforwardly. Therefore, we compare our result with the best two methods for the English language in the CLEF IP 2011 competition based on Figure 3 from \citet{piroi2011clef}: Ch.2 and Hy.5. We found that Ch.2 is the most effective method on CLEF-IP 2011, and thus we used that as our initial ranker.
\par
For the Wikipedia datasets, we compare our result with $SDR_{inf}$ which proposed these datasets recently and is the state-of-the-art model for them.
\subsection{Pre-trained Sentence BERT models}\label{sec:pretrained_models}
\citet{reimers2019sentence} have published a set of SBERT embedding models that they trained and evaluated extensively with respect to their effectiveness for semantic textual similarity (Sentence Embeddings) and Semantic Search tasks on 14 and 5 different datasets respectively.\footnote{\href{https://www.sbert.net/docs/pretrained_models.html}{https://www.sbert.net/docs/pretrained\_models.html}} We exploit the following top-$4$ ranked embedding models\footnote{Please note that these models are top-ranked at the time of submission and the SBERT list can be updated later.} that are different in terms of training data or architecture:
\begin{itemize}
    \item \textbf{all-mpnet-base-v2} \footnote{https://huggingface.co/sentence-transformers/all-mpnet-base-v2}, \textbf{all-distilroberta-v1} \footnote{https://huggingface.co/sentence-transformers/all-distilroberta-v1}, and \textbf{all-MiniLM-L12-v2} \footnote{https://huggingface.co/sentence-transformers/all-MiniLM-L12-v2} are SBERT models trained based on \emph{mpnet-base} \cite{song2020mpnet}, \emph{distilbert-base-cased} \cite{sanh2019distilbert} and \emph{MiniLM-L12-H384-uncased} \cite{wang2020minilm} models respectively on more than $1$ billion sentence pairs as general purpose models for sentence similarity;
    \item \textbf{multi-qa-mpnet-base-dot-v1} \footnote{https://huggingface.co/sentence-transformers/multi-qa-mpnet-base-dot-v1} is a \emph{mpnet-base} model \cite{song2020mpnet} trained on $215M$ question-answer pairs from various sources and domains, including StackExchange, Yahoo Answers, Google and Bing search queries and many more as a model for question answering and IR tasks.
\end{itemize}
Besides the above mentioned SBERT models, we utilize the following domain specific and general well-known variants of BERT and Roberta models in the architecture of SBERT to investigate their effectiveness,
considering that they have the efficiency of the SBERT's architecture: 
\begin{itemize}
    \item BERT base uncased, BERT large uncased, Roberta base uncased, and Roberta large uncased are pre-trained on a large corpus of English raw data  \cite{devlin2018bert,liu2019roberta};%
    \item Legal BERT base uncased \cite{chalkidis2020legal} is 
    a 
    light-weight model of BERT-BASE
    (33\% the size of BERT-BASE) 
    pre-trained from scratch on $12$ GB of diverse English legal text of several types (e.g., legislation, court cases, contracts);
    \item Patent BERT \cite{patentBERT} is trained by Google on $100M+$ patents (not just US patents) including abstract, claims, description based on BERT large uncased architecture.
\end{itemize}
We only report the results of the trained SBERT models as they  obtained higher than the general well-known variants of BERT models in the architecture of SBERT in all datasets except for the Wikipedia datasets (WWG and WWA). For these, we report the Roberta large model in the architecture of SBERT, which achieved the best result.
\subsection{Implementation details}\label{sec:implementation}
For SBERT, we first experiment with all the models that are introduced in section \ref{sec:pretrained_models} and then adapting the two best models for each dataset on the domain using the methods proposed by \citet{wang2021tsdae} (TSDAE) and \citet{ginzburg-etal-2021-self} (SDR) for domain adaptation without labeled data.\footnote{The resulting models on the patent, legal and Wikipedia domain will be publicly available.}
The SDR pre-training method is a dual-term loss objective that is composed of a standard MLM loss adopted from \cite{devlin2018bert} -- which allows the model to specialize in the domain of the given collection \cite{gururangan2020don} -- and a contrastive loss \cite{hadsell2006dimensionality}. For SDR pre-training, we tokenize sentences, aggregate them into batches, and randomly mask them in a similar way to the RoBERTa pre-training paradigm. %
For TSDAE pre-training, as suggested by \cite{wang2021tsdae}, we use $10,000$ sentences for each domain which is only $2\%$ of the COLIEE'21, $1\%$ of the Wikipedia (WWG and WWA) datasets, and less than $0.1\%$ of the CLEF-IP 2011 dataset. PyTorch \cite{paszke2019pytorch}, HuggingFace \cite{wolf2019huggingface}, and Sentence BERT \cite{reimers2019sentence} are used to implement
all of our models.
\par
We use $BM25_{optimised}$ \cite{askari2021combining} as our initial ranker for COLIEE, \emph{K} for Caselaw \cite{locke2017automatic}, and \emph{Ch.2} for CLEF-IP 2011 \citep{piroi2011clef}.
To determine the optimal re-ranking depth, we increase the depth of the initial rank result on the validation set from $15$ to $100$ in steps of $5$. We found $50$ to be optimal for COLIEE, $100$ for WWG and WWA, and $20$ for the CLEF-IP 2011 dataset.
As we use Caselaw as test set only, we employ COLIEE's re-ranking depth for Caselaw.

\par
\subsection{Parameter tuning}\label{sec:parameter_tuning}
For tuning parameters we experiment with values from $1$ to $10$ for parameter $n$: $n=[1,2,3,4,5,6,7,8,9,10]$. For the $b$ and $k1$ parameters of \emph{RPRS w/freq}, we used the same range as when tuning them for BM25 ($b=[0.0,0.1,0.2,\cdots,1], k1=[0.0,0.2,0.4,\cdots,3.0]$). We found $(b=1, k1=2.8, n=4)$ for COLIEE'21, $(b=0.9, k1=3.0, n=4)$ for WWG and WWA, and $(b=0.8, k1=2.4, n=5)$ for CLEF-IP.
\subsection{Pre-processing}
We investigated sentence segmentation using three libraries since legal case retrieval and patent retrieval are challenging domain-specific tasks with long legal sentences  \cite{sanchez-2019-sentence,verberne2010quantifying}. We empirically found that SparkNLP \cite{KOCAMAN2021100058} is segmenting the legal text into sentences with a higher quality than NLTK \cite{bird2009natural} and Stanza \cite{qi2020stanza}.
We split long sentences into sequences of $25$ words for COLIEE and $30$ words for CLEF-IP (average length of sentences in both collections). We then detected French sentences in the COLIEE data using the method by \cite{danilak2014langdetect} and translated them to English using the Google Translate API. %

\section{Results} \label{sec:results}
We employ a paired t-test between the proposed method, i.e., \emph{RPRS w/freq}, and the state-of-the-art model for each dataset. 
We implement four mutual baselines (BM25, BM25+KLI, SDR\textsubscript{inf}, Birch) across all the datasets. Additionally, we present state-of-the-art approaches for each individual dataset. By doing so, we guarantee the comprehensive coverage of state-of-the-art methodologies for each dataset. It is worth highlighting that, to the best of our knowledge, no previous work have undertaken the challenge of addressing query-by-document tasks across all these datasets collectively. Existing works have been limited to individual datasets, each concentrating on a single dataset.
\par
We answer the following research questions, assessing the effectiveness of our proposed methods, RPRS and RPRS w/freq, from different perspectives:
\begin{itemize}
    \item \textbf{RQ1: }What is the effectiveness of RPRS compared to the State-of-the-art models for QBD retrieval?
    \item \textbf{RQ2: }How effective is RPRS with shorter or longer text units instead of sentences?
    \item \textbf{RQ3: }What is the effectiveness of RPRS with parameters that were tuned on a different dataset in the same domain?
    \item \textbf{RQ4: }To what extent is RPRS effective and generalizable across different domains with different type of documents?
\end{itemize}
\par
In the following, we first address the choice of SBERT model for the proposed method, \emph{RPRS w/freq}. Next, we analyze the effectiveness of the proposed method on the five domain-specific QBD datasets and discuss the results per domain. In summary, we found that \emph{RPRS w/freq} outperforms the state-of-the-art models significantly for all official metrics in all five datasets. 
Please note that the initial ranker remains consistent across various re-rankers for all datasets, thereby establishing a fair comparison. Furthermore, each of the BERT-based re-rankers utilizes the identical BERT model employed by RPRS. This choice ensures that RPRS does not gain an unfair advantage from using a potentially superior BERT model compared to the baseline re-rankers.%
\begin{table}
\centering
\caption{Results of the Proposed Method, \emph{RPRS w/freq}, on COLIEE'21 using different Transformer Models Embedding. Top SBERT models are top-$4$ trained and extensively evaluated models by SBERT and are publicly available. Legal SBERT is the Legal BERT \cite{chalkidis2020legal} that is loaded into the SBERT \cite{reimers2019sentence} architecture. Domain adaption has been done using Self-Supervised, SDR, and Unsupervised, TSDAE, methods on all-MiniLM-L12-v2 and Legal SBERT models as the models show top-two highest effectiveness without Domain Adaption.}
\label{tab:optimal_representation}
\arrayrulecolor[rgb]{0.753,0.753,0.753}
\begin{tabular}{lc|c|c|c} 
\arrayrulecolor{black}\hline
 \rule[-2ex]{0pt}{0pt}\rule{0pt}{3ex}& Model Name & Precision & Recall & F1 \\ 
\arrayrulecolor[rgb]{0.502,0.502,0.502}\hline
\multicolumn{2}{l!{\color{silver}\vrule}}{\rule[-2ex]{0pt}{0pt}\rule{0pt}{3ex}\textbf{Initial Ranker}} & \multicolumn{1}{l!{\color{silver}\vrule}}{} & \multicolumn{1}{l!{\color{silver}\vrule}}{} & \multicolumn{1}{l}{} \\
\rule[-2ex]{0pt}{0pt}a & BM25\textsubscript{optimized} + KLI & 0.1700 & 0.2536 & 0.2035 \\ \hline
\multicolumn{2}{l!{\color{silver}\vrule}}{\rule[-2ex]{0pt}{0pt}\rule{0pt}{3ex}\textbf{Top SBERT models}} & \multicolumn{1}{l!{\color{silver}\vrule}}{} & \multicolumn{1}{l!{\color{silver}\vrule}}{} & \multicolumn{1}{l}{} \\ 
b & all-mpnet-base-v2 & 0.1760 & 0.2478 & 0.2058 \\
c & multi-qa-mpnet-base-dot-v1 & 0.1840 & 0.2625 & 0.2163 \\
d & all-distilroberta-v1 & 0.1900 & 0.2632 & 0.2206 \\
e & all-MiniLM-L12-v2 & 0.1920 & 0.2745 & 0.2259 \\
\hline
\multicolumn{2}{l!{\color{silver}\vrule}}{\rule[-2ex]{0pt}{0pt}\rule{0pt}{3ex}\textbf{Domain-Specific}} & \multicolumn{1}{l!{\color{silver}\vrule}}{} & \multicolumn{1}{l!{\color{silver}\vrule}}{} & \multicolumn{1}{l}{} \\ 

f & Legal SBERT & 0.1880 & 0.2720 & 0.2223 \\ 
\hline
\multicolumn{2}{l!{\color{silver}\vrule}}{\rule[-2ex]{0pt}{0pt}\rule{0pt}{3ex}\textbf{Domain Adaptation}} & \multicolumn{1}{l!{\color{silver}\vrule}}{} & \multicolumn{1}{l!{\color{silver}\vrule}}{} & \multicolumn{1}{l}{} \\
g & all-MiniLM-L12-v2-SDR & 0.1817 & 0.2594 & 0.2137 \\
h & all-MiniLM-L12-v2-TSDAE & 0.1840 & 0.2584 & 0.2149 \\
i & Legal SBERT-SDR & 0.1890 & 0.2762 & 0.2244 \\
\rule[-2ex]{0pt}{0pt}j & Legal SBERT-TSDAE & \textbf{0.1960} & \textbf{0.2891} & \textbf{0.2336} \\
\arrayrulecolor{black}\hline
\end{tabular}
\end{table}
\subsection{Choice of SBERT model for \emph{RPRS w/freq}}
In order to address \textbf{RQ1}, we first run  a set of experiments with different SBERT models to find with which SBERT model the \emph{RPRS w/freq} achieve higher effectiveness. Table \ref{tab:optimal_representation} shows that our re-ranker achieves a higher effectiveness than the strong initial ranker (line $a$) \cite{askari2021combining} using any sentence embedding models (line $b-j$). This shows the effectiveness of our re-ranker is more dependent on the proposed method rather than sentence embedding models, while our re-ranker effectiveness could be improved with a sentence embedding model that captures the legal context more accurate. Morover, the table shows that out of the four SBERT models (line $b-e$),  \emph{all-MininLM-L12-v2}, achieves highest effectiveness. Interestingly, this is the SBERT model with the lowest number of parameters ($33M$, compared to $82M$, $110M$, and $110M$ for the other three). This could indicate that between SBERT models which have shown their effectiveness in other (general) domains, the chance of generalizing to a new domain-specific dataset for a SBERT model that has lowest parameters might be higher than others. This finding is in line with the fact that, in the context of traditional bias-variance trade-offs, neural models with fewer parameters are more likely to generalize to the other domains \cite{zhong2021larger,zhou2021rethinking}. The higher effectiveness for \emph{multi-qa-mpnet-base-dot-v1} (line $c$) compared to \emph{all-mpnet-base-v2} (line $b$) while they both have same number of parameters and same base architecture could be due to the fact that \emph{multi-qa-mpnet-base-dot-v1} was trained on (question, pair) instances, and as a result, it is suitable for semantic search task and aligns with our re-ranking task while all-mpnet-base-v2 is trained on various tasks.%
\par
\begin{table}[t]
\centering
\caption{Results for COLIEE’21.$\dagger$ and * indicate a statistically significant improvement over State-of-the-art (MTFT-BERT, line g) and BM25\textsubscript{optimized} (line c) respectively, according to a paired-test (p<0.005) with Bonferroni correction for multiple testing. The winning team in the COLIEE 2021 competition for legal case retrieval is TLIR \cite{ma2021retrieving}. The initial ranker for all re-rankers (BERT, MTFT \cite{askari2021combining}, SDR, and RPRS ) is BM25\textsubscript{optimized}+KLI. Legal SBERT-TSDAE is used to embed sentences for $SDR_{inf}$, RPRS, and \emph{RPRS w/freq}.}
\label{tab:res_coliee}
\arrayrulecolor[rgb]{0.753,0.753,0.753}
\begin{tabular}{lll|c|c|c} 
\arrayrulecolor{black}\hline
 \rule[-2ex]{0pt}{0pt}\rule{0pt}{3ex}& & Model & Precision & Recall & F1 \\ 
\arrayrulecolor[rgb]{0.502,0.502,0.502}\hline
 \rule[-2ex]{0pt}{0pt}\rule{0pt}{3ex}& \multicolumn{2}{l!{\color{silver}\vrule}}{\textbf{Probabilistic lexical matching baselines}} & \multicolumn{1}{l!{\color{silver}\vrule}}{} & \multicolumn{1}{l!{\color{silver}\vrule}}{} & \multicolumn{1}{l}{} \\
 & a & BM25 & 0.0770 & 0.1959 & 0.1113 \\
 & b & BM25 + KLI & 0.0983 & 0.1980 & 0.1313 \\
 & c & BM25optimized + KLI & 0.1700 & 0.2536 & 0.2035 \\
 \rule[-2ex]{0pt}{0pt}& d & TLIR & 0.1533 & 0.2556 & 0.1917 \\ 
\hline
 \rule[-2ex]{0pt}{0pt}\rule{0pt}{3ex}& \multicolumn{2}{l!{\color{silver}\vrule}}{\textbf{Cross-encoders}} & \multicolumn{1}{l!{\color{silver}\vrule}}{} & \multicolumn{1}{l!{\color{silver}\vrule}}{} & \multicolumn{1}{l}{} \\
 & e & BERT & 0.1340 & 0.2263 & 0.1683 \\
 & f & Legal BERT & 0.1440 & 0.2463 & 0.1817 \\
 \rule[-2ex]{0pt}{0pt}& g & \multicolumn{1}{l!{\color{silver}\vrule}}{MTFT-BERT \cite{AbolghasemiECIR22} (Previous state-of-the-art)} & 0.1744 & \textbf{0.2999} & 0.2205 \\ 
\hline
 \rule[-2ex]{0pt}{0pt}\rule{0pt}{3ex}& \multicolumn{2}{l!{\color{silver}\vrule}}{\textbf{Sentence-based baseline}} & \multicolumn{1}{l!{\color{silver}\vrule}}{} & \multicolumn{1}{l!{\color{silver}\vrule}}{}  & \multicolumn{1}{l}{} \\ 
\arrayrulecolor{silver}
 \rule[-2ex]{0pt}{0pt}& h & $SDR_{inf}$ & 0.1470 & 0.2063 & 0.1716 \\ 
  & i & Birch & 0.1721 & 0.2577 & 0.2064 \\
\arrayrulecolor[rgb]{0.502,0.502,0.502}\hline
 \rule[-2ex]{0pt}{0pt}\rule{0pt}{3ex}& \multicolumn{2}{l!{\color{silver}\vrule}}{\textbf{\textbf{\textbf{\textbf{Proposed methods}}}}} & \multicolumn{1}{l!{\color{silver}\vrule}}{} & \multicolumn{1}{l!{\color{silver}\vrule}}{} & \multicolumn{1}{l}{} \\
 & j & RPRS without frequency & 0.1890 & 0.2799 & 0.2256 \\
 \rule[-2ex]{0pt}{0pt}& k & RPRS w/freq & \textbf{\textbf{0.1960$\dagger$*}} & 0.2891* & \textbf{\textbf{0.2336$\dagger$*}} \\
\arrayrulecolor{black}\hline
\end{tabular}
\end{table}
The second highest effectiveness is achieved by \emph{Legal SBERT}, which is the domain-specific Legal BERT model \cite{chalkidis2020legal} that is loaded into the SBERT architecture \cite{reimers2019sentence} and has $36M$ parameters (more than MiniLM and less than our three other SBERT models). The effectiveness of our proposed method with the usage of Legal SBERT model could be due the fact it is pre-trained on the legal documents. However, with domain adaptation on COLIEE corpus using TSDAE method \cite{wang2021tsdae}, i.e., Legal SBERT-TSDAE, \emph{RPRS w/freq} achieves the highest effectiveness (row $j$). We found the same pattern for SDR and RPRS without frequency (section \ref{sec:proposed_method_pb}), i.e., they achieve highest effectiveness with Legal-BERT-TSDAE model embeddings, but we only report it for \emph{RPRS w/freq} which is our complete proposed method.
\par
Furthermore, an observation by comparing rows ($g,i$) and ($h,j$) of Table \ref{tab:optimal_representation}, is that the domain adaption with TSDA results in higher effectiveness of our re-ranker compared to SDR domain adaption method which is a contrastive-learning domain adaption method. We suppose the higher effectiveness of our re-ranker with domain adaption by TSDAE compared to SDR could be due to the fact that TSDAE removes words from the input randomly and as the result forces the network to produce robust embeddings. In contrast, the input sentences for a contrastive-learning approach such as SDR are not modified, resulting in less stable embeddings. This finding was confirmed previously by Wang et al. \cite{wang2021tsdae}.
\par
Additionally, by comparing rows $e$ with $g$ and $h$, we can see that domain adaption on all-MiniLM-L12-v2 results in drops in the effectiveness of our re-ranker. This observation is in line with the finding in prior work \cite{wang2021tsdae} which shows that domain adaption with TSDAE after supervised training results in a drop in the effectiveness. It is noteworthy to mention that all-MiniLM-L12-v2 is trained on 1 billion samples in a supervised manner. A better strategy in this case, would be first doing domain adaption as a pre-training step with TSDAE and then doing supervised training with 1 billion samples as was suggested by the authors of TSDAE \cite{wang2021tsdae}.\footnote{Also reported on \url{https://github.com/UKPLab/sentence-transformers/issues/1372\#issuecomment-1028525300} (visited February 27, 2023)} However, re-training MiniLM with 1 billion examples was not possible for us due to computational limitations.
\par
Therefore, based on our empirical findings, for the next experiments on each domain, we pick the domain-specific Transformer model and then adapt it into the the target domain, e.g., legal, using TSDAE. We argue that doing this domain adaption is realistic for a real-world application because it is trained on the document collection that are provided for training, and does not need or use any supervision from labelled data.
\subsection{Effectiveness on the COLIEE dataset (\textbf{RQ1})}
Taking into account the finding from previous section, we exploit Legal SBERT-TSDAE in Table \ref{tab:res_coliee} to embed sentences for \emph{RPRS}, \emph{RPRS w/freq}, and SDR. Row $i$ of Table \ref{tab:res_coliee} show that even without taking into account frequency, \emph{RPRS} achieves higher effectiveness than the initial ranker ($BM25_{optimized}+KLI$, line $c$), Birch (line $i$), and the state-of-the-art re-ranker (MTFT-BERT, line $g$) on precision and the official metric ($F1$). This is while \emph{RPRS} has only one parameter, $n$. 
Row $k$ of the table shows that $RPRS w/freq$ achieve highest effectiveness and significantly better results over the state-of-the-art (MTFT-BERT\cite{AbolghasemiECIR22}, line $g$) on precision and the official metric ($F1$) which indicates the importance and role of adding the document length normalization parameter $b$, and frequency saturation parameter $k1$ in increasing the effectiveness of the proposed method. 
We emphasize that the official metric for COLIEE is $F1$ and the COLIEE organizers only report $F1$ in their official reports for case law retrieval task \cite{goebel2021coliee}. We assume achieving lower recall by our re-ranker, \emph{RPRS w/freq} (line $k$), compared to MTFT-BERT \cite{AbolghasemiECIR22} (line $g$) could be due to the inverse relationship between recall and precision which is a common observation based on experimental evidence \cite{buckland1994relationship}. \citet{cleverdon1972inverse} provides an in-depth explanation of why precision and recall have often an inverse relationshi. $SDR_{inf}$ (line $h$) achieves lower performance than $RPRS w/freq$ and $RPRS$ results (line j and k).
\begin{figure}[t]
\scriptsize
\centering
\scalebox{.7}{\includegraphics[]{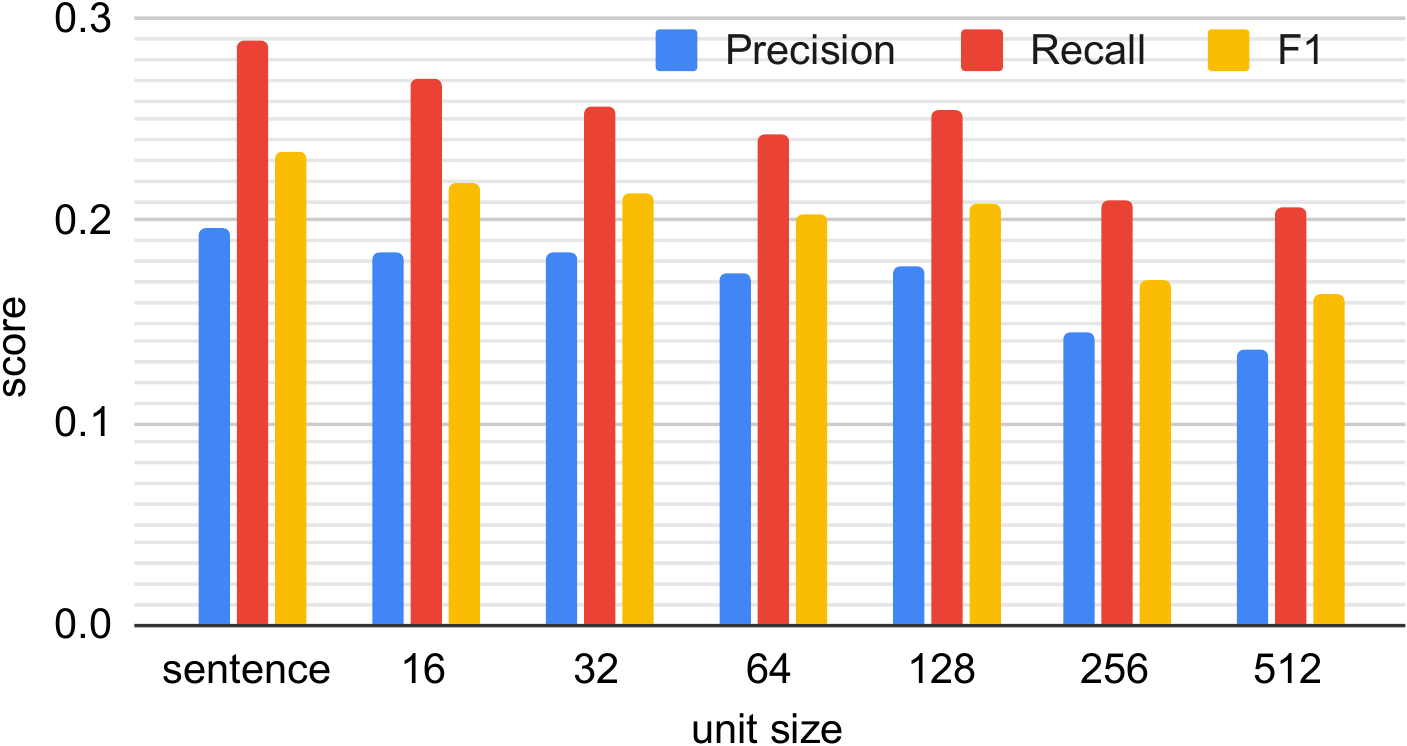}}
\caption{Effectiveness of RPRS w/freq using fixed-length units instead of sentences on COLIEE 2021 (segment size in tokens).}%
\label{fig:analysis_unit_effect}
\end{figure}
\subsection{Effect of using other units than sentences \textbf{(RQ2)}}
We investigate the effect of using embeddings of different textual units than sentences in \emph{RPRS w/freq} to analyze if ``sentence'' is the most appropriate unit for our re-ranker. 
To do so, we split the documents in segments of length $l$ with $l \in \{16,32,64,128,256,512\}$. Figure \ref{fig:analysis_unit_effect} shows that sentences are better units than sequences with pre-defined fixed-length, also compared to sequences
with fixed-length $16$ and $32$
which are similar in length to the average and median sentence length ($29.4$ and $24.8$ tokens). This confirms the relevance of sentences as units for retrieval, one of the premises of RPRS.
\begin{table}
\centering
\caption{Results for CaseLaw. $\dagger$ indicate a statistically significant improvement over the state-of-the-art (K \cite{locke2017automatic}, line c) according to a paired-test ($p<0.05$) with Bonferroni correction for multiple testing. We pick the optimized values of the proposed method parameters from COLIEE dataset, and use them to analyze the proposed method cross-data generalizability on Caselaw dataset. Legal SBERT-TSDAE is used to embed sentences for $SDR_{inf}$, RPRS, and \emph{RPRS w/freq}.}
\label{tab:res_caselaw}
\arrayrulecolor[rgb]{0.753,0.753,0.753}
\begin{tabular}{ll|c|c|c|c|c} 
\arrayrulecolor{black}\hline
 \rule[-2ex]{0pt}{0pt}\rule{0pt}{3ex}& Model & P@1 & R@1 & MAP@5 & NDCG@5 & MRR \\ 
\hline
\multicolumn{2}{l!{\color{silver}\vrule}}{\rule[-2ex]{0pt}{0pt}\rule{0pt}{3ex}\textbf{Probabilistic lexical matching baselines}} & \multicolumn{1}{l!{\color{silver}\vrule}}{} & \multicolumn{1}{l!{\color{silver}\vrule}}{} & \multicolumn{1}{l!{\color{silver}\vrule}}{} & \multicolumn{1}{l!{\color{silver}\vrule}}{} & \multicolumn{1}{l}{} \\
a & BM25 & 0.500 & 0.025 & 0.227 & 0.358 & 0.618 \\
b & BM25 + KLI & 0.660 & 0.115 & 0.232 & 0.398 & 0.726 \\
\rule[-2ex]{0pt}{0pt}c & K \cite{locke2017automatic} (Previous State-of-the-art) & 0.730 & 0.119 & 0.307 & 0.473 & 0.803 \\ 
\arrayrulecolor[rgb]{0.502,0.502,0.502}\hline
\multicolumn{2}{l!{\color{silver}\vrule}}{\rule[-2ex]{0pt}{0pt}\rule{0pt}{3ex}\textbf{Cross-encoders}} & \multicolumn{1}{l!{\color{silver}\vrule}}{} & \multicolumn{1}{l!{\color{silver}\vrule}}{} & \multicolumn{1}{l!{\color{silver}\vrule}}{} & \multicolumn{1}{l!{\color{silver}\vrule}}{} & \multicolumn{1}{l}{} \\
d & BERT & 0.320 & 0.064 & 0.118 & 0.201 & 0.438 \\
e & Legal BERT & 0.330 & 0.065 & 0.128 & 0.217 & 0.449 \\
\rule[-2ex]{0pt}{0pt}f & MTFT-BERT & 0.360 & 0.064 & 0.160 & 0.252 & 0.481 \\ 
\hline
\multicolumn{2}{l!{\color{silver}\vrule}}{\rule[-2ex]{0pt}{0pt}\rule{0pt}{3ex}\textbf{Sentence-based baseline}} & \multicolumn{1}{l!{\color{silver}\vrule}}{} & \multicolumn{1}{l!{\color{silver}\vrule}}{} & \multicolumn{1}{l!{\color{silver}\vrule}}{} & \multicolumn{1}{l!{\color{silver}\vrule}}{} & \multicolumn{1}{l}{} \\
\rule[-2ex]{0pt}{0pt}g & $SDR_{inf}$ & 0.500 & 0.070 & 0.171 & 0.320 & 0.585 \\ 
\rule[-2ex]{0pt}{0pt}h & Birch & 0.680 & 0.117 & 0.270 & 0.425 & 0.774 \\ 
\hline
\multicolumn{2}{l!{\color{silver}\vrule}}{\rule[-2ex]{0pt}{0pt}\rule{0pt}{3ex}\textbf{Proposed methods}} & \multicolumn{1}{l!{\color{silver}\vrule}}{} & \multicolumn{1}{l!{\color{silver}\vrule}}{} & \multicolumn{1}{l!{\color{silver}\vrule}}{} & \multicolumn{1}{l!{\color{silver}\vrule}}{} & \multicolumn{1}{l}{} \\
i & RPRS without frequency & 0.730 & 0.128 & 0.314 & 0.489 & 0.807 \\
\rule[-2ex]{0pt}{0pt}j & RPRS w/freq & \textbf{0.780$\dagger$} & \textbf{0.138$\dagger$} & \textbf{0.321$\dagger$} & \textbf{0.496$\dagger$} & \textbf{0.837$\dagger$} \\
\arrayrulecolor{black}\hline
\end{tabular}
\end{table}
\subsection{Effectiveness without parameter optimization on target dataset\textbf{(RQ3)}}
To address this question, we evaluate the proposed method on the \emph{Caselaw} dataset \cite{locke2017automatic} with the parameters that were tuned on the COLIEE dataset without doing domain adaption on the collection for the Transformer model. In other words, our goal is to analyze how much the three optimized parameters of RPRS on the COLIEE dataset are transferable to another dataset in the same (i.e., legal) domain.
Table \ref{tab:res_caselaw} shows that the effectiveness of our re-ranker is higher than all models including the state-of-the-art model (\emph{K}, line $c$) for all evaluation metrics with (line $i$) or without (line $j$) taking into account frequency.
This supports the cross-data generalizability of our re-ranker, i.e., optimizied parameters of our proposed method work effectively for another dataset in the same domain, as we re-use the tuned $RPRS w/freq$ on COLIEE for the Caselaw dataset without optimizing its parameters on the Caselaw dataset. $SDR_{inf}$ (line $g$) and Birch (line $h$) baselines achieve lower performance than $RPRS w/freq$ and $RPRS$ results (line $i-j$).
\subsection{Generalizability of porposed method (RQ4)}\label{sec:generalizability}
\textbf{Patent domain (CLEF-IP 2011 dataset).}
Table \ref{tab:res_patent} shows that our re-ranker outperform the state-of-the-art model on the CLEF-IP 2011 for all evaluation metrics with or without taking into account frequency (lines $g$ and $h$). 
Row $g$ of Table \ref{tab:res_patent} show that even without taking into account frequency \emph{RPRS} achieves higher effectiveness than the initial ranker (Ch.2, line $d$), the previous state-of-the-art on CLEF-IP'11. Patent SBERT-TSDAE is used to embed sentences for $SDR_{inf}$, RPRS, and \emph{RPRS w/freq}, which is Patent BERT \cite{patentBERT} model that is loaded into into the SBERT \cite{reimers2019sentence} architecture and adapted into the domain using TSDAE method \cite{wang2021tsdae}. $SDR_{inf}$ (line $e$) and Birch (line $f$) show competitive results compared to BM25 (line $a$ and $b$), however, it obtained much lower results than $RPRS w/freq$. 

\begin{table}
\centering
\caption{Results for Patent Retrieval. $\dagger$ indicate the statistically significant improvement over best team on the CLEF-IP11 data (Ch.2 \cite{piroi2011clef}) according to a paired-test (p<0.001) with Bonferroni correction for multiple testing respectively. Patent SBERT-TSDAE is used to embed sentences for $SDR_{inf}$, RPRS, and \emph{RPRS w/freq}.}
\label{tab:res_patent}
\resizebox{15cm}{!}{
    \arrayrulecolor[rgb]{0.753,0.753,0.753}
    \begin{tabular}{ll|c|c|c|c|c|c|l} 
    \arrayrulecolor{black}\hline
     & \rule[-2ex]{0pt}{0pt}\rule{0pt}{3ex}Model & P@5 & P@10 & R@5 & R@10 & MAP@5 & MAP@10 & MRR \\ 
    \arrayrulecolor[rgb]{0.502,0.502,0.502}\hline
    \multicolumn{2}{l!{\color{silver}\vrule}}{\rule[-2ex]{0pt}{0pt}\rule{0pt}{3ex}\textbf{Probabilistic lexical matching baselines}} & \multicolumn{1}{l!{\color{silver}\vrule}}{} & \multicolumn{1}{l!{\color{silver}\vrule}}{} & \multicolumn{1}{l!{\color{silver}\vrule}}{} & \multicolumn{1}{l!{\color{silver}\vrule}}{} & \multicolumn{1}{l!{\color{silver}\vrule}}{} & \multicolumn{1}{l!{\color{silver}\vrule}}{} & \\
    a & BM25 & 0.084 & 0.066 & 0.010 & 0.147 & 0.053 & 0.065 & 0.187 \\
    \rule[-2ex]{0pt}{0pt}b & BM25 + KLI & 0.095 & 0.071 & 0.011 & 0.167 & 0.074 & 0.087 & 0.221 \\ 
    \hline
    \multicolumn{2}{l!{\color{silver}\vrule}}{\textbf{\rule[-2ex]{0pt}{0pt}\rule{0pt}{3ex}Top-two teams in English CLEF-IP'11}} & \multicolumn{1}{l!{\color{silver}\vrule}}{} & \multicolumn{1}{l!{\color{silver}\vrule}}{} & \multicolumn{1}{l!{\color{silver}\vrule}}{} & \multicolumn{1}{l!{\color{silver}\vrule}}{} & \multicolumn{1}{l!{\color{silver}\vrule}}{} & \multicolumn{1}{l!{\color{silver}\vrule}}{} & \\
    c & Hy.5  & 0.057 & 0.041 & 0.071 & 0.103 & 0.045 & 0.051 & 0.162 \\
    \rule[-2ex]{0pt}{0pt}d & Ch.2 \cite{piroi2011clef} (Previous state-of-the-art) & 0.122 & 0.089 & 0.150 & 0.216 & 0.102 & 0.118 & 0.296 \\ 
    \hline
    \multicolumn{2}{l!{\color{silver}\vrule}}{\rule[-2ex]{0pt}{0pt}\rule{0pt}{3ex}\textbf{Sentence-based baseline}} & \multicolumn{1}{l!{\color{silver}\vrule}}{} & \multicolumn{1}{l!{\color{silver}\vrule}}{} & \multicolumn{1}{l!{\color{silver}\vrule}}{} & \multicolumn{1}{l!{\color{silver}\vrule}}{} & \multicolumn{1}{l!{\color{silver}\vrule}}{} & \multicolumn{1}{l!{\color{silver}\vrule}}{} & \\
    \rule[-2ex]{0pt}{0pt}e & $SDR_{inf}$ + Ch.2 & 0.105 & 0.077 & 0.120 & 0.175 & 0.078 & 0.090 & 0.251 \\
    \rule[-2ex]{0pt}{0pt}f & Birch + Ch.2 & 0.125 & 0.087 & 0.148 & 0.208 & 0.104 & 0.114 & 0.288 \\
    \hline
    \multicolumn{2}{l!{\color{silver}\vrule}}{\rule[-2ex]{0pt}{0pt}\rule{0pt}{3ex}\textbf{Proposed methods}} & \multicolumn{1}{l!{\color{silver}\vrule}}{} & \multicolumn{1}{l!{\color{silver}\vrule}}{} & \multicolumn{1}{l!{\color{silver}\vrule}}{} & \multicolumn{1}{l!{\color{silver}\vrule}}{} & \multicolumn{1}{l!{\color{silver}\vrule}}{} & \multicolumn{1}{l!{\color{silver}\vrule}}{} & \\
    g & RPRS + Ch.2 & 0.131 & 0.092 & 0.160 & 0.221 & 0.113 & 0.129 & 0.319 \\
    \rule[-2ex]{0pt}{0pt}h & RPRS w/freq + Ch.2 & \textbf{0.132$\dagger$} & \textbf{0.093$\dagger$} & \textbf{0.167$\dagger$} & \textbf{0.229$\dagger$} & \textbf{0.116$\dagger$} & \textbf{0.132$\dagger$} & \textbf{0.332$\dagger$} \\ 
    \arrayrulecolor{black}\hline
    \end{tabular}
}
\end{table}

\noindent \textbf{Wikipeda domain (Wine and Video games datasets).}
Table \ref{tab:sdr_datasets_results} shows that $SDR_{inf}$ (line $c$ and $d$) and Birch (line $e$) obtain competitive results to $BM25+KLI$ (line $b$) on SDR's own datasets. However, as we achieved higher recall with BM25+KLI, we use it as the initial ranker for our re-ranker.
Results show the generalizability and effectiveness of $RPRS w/freq$ and $RPRS$ on a very different domain (Wikipedia) for two datasets as they could outperform the state-of-the-art model on the WWG and WWA datasets for all evaluation metrics with or without taking into account frequency (lines $f$ and $g$). It is noteworthy to mention that we report all of the official metrics of WWG and WWA datasets. We refer to \emph{Hit Ratio $@k$} in the original paper \cite{ginzburg-etal-2021-self} as recall in this table as their definition for the \emph{Hit Ratio} is equal to recall in Information Retrieval. We compute Mean Percentile Rank ($MPR$) using the original implementation by SDR's authors \cite{ginzburg-etal-2021-self}.\footnote{https://github.com/microsoft/SDR/blob/main/models/reco/wiki\_recos\_eval/eval\_metrics.py}
\begin{table}[ht]
\caption{Results for the video games (left), and wines (right) datasets from~\citet{ginzburg-etal-2021-self}. R referse to Roberta large in this table. $\dagger$ and * indicate a statistically significant improvement over previous State-of-the-art (SDR, line b) and BM25+KLI (line d) respectively, according to a paired-test ($p<0.005$) with Bonferroni correction for multiple testing.}
\label{tab:sdr_datasets_results}
\resizebox{15cm}{!}{
    \begin{tabular}{llllcccccccc}
    \cline{1-7} \cline{9-12} \multirow{2}{*}{} & \multirow{2}{*}{\rule[-2ex]{0pt}{0pt}\rule{0pt}{3ex}Model} & \multirow{2}{*}{Embedding} & \multicolumn{4}{c}{\rule[-2ex]{0pt}{0pt}\rule{0pt}{3ex}Video games (WWG)} & & \multicolumn{4}{c}{Wines (WWA)} \\ \cline{4-7} \cline{9-12} 
    \rule[-2ex]{0pt}{0pt}\rule{0pt}{3ex}& & & recall@10 & recall@100 & MPR & MRR & & recall@10 & recall@100 & MPR & MRR \\ \hline
    \multicolumn{12}{l}{\textbf{\rule[-2ex]{0pt}{0pt}\rule{0pt}{3ex}Probabilistic lexical matching baselines}} \\
    a & BM25 & - & 0.2200 & 0.4887 & 0.8606 & 0.5681 & & 0.1704 & 0.5041 & 0.8164 & 0.4562 \\
    \rule[-2ex]{0pt}{0pt}b & BM25 + KLI & - & 0.2425 & \textbf{0.5499} & 0.8748 & 0.6150 & & 0.1732 & \textbf{0.6332} & 0.8191  & 0.4759 \\ \hline
    \multicolumn{12}{l}{\textbf{\rule[-2ex]{0pt}{0pt}\rule{0pt}{3ex}Previous state-of-the-art}} \\
    c & $SDR_{inf}$  & R SDR & 0.2360 & 0.5400 & 0.9740 & 0.6400 & & 0.1700 & 0.5900 & 0.8930 & 0.5090 \\
    \rule[-2ex]{0pt}{0pt}d & $SDR_{inf}$  & R TSDAE  & 0.2384 & 0.5425 & 0.9761 & 0.6433 & & 0.1712 & 0.5980 & 0.8940 & 0.5100 \\ \hline
    \multicolumn{12}{l}{\textbf{\rule[-2ex]{0pt}{0pt}\rule{0pt}{3ex}Sentence-BERT baseline}} \\
    \rule[-2ex]{0pt}{0pt}e & Birch & R TSDAE  & 0.2322 & 0.5271 & 0.8683 & 0.6115 & & 0.1720 & 0.5599 & 0.8088 & 0.4680 \\ \hline
    \multicolumn{12}{l}{\textbf{\rule[-2ex]{0pt}{0pt}\rule{0pt}{3ex}Proposed Methods}} \\
    f & RPRS  & R TSDAE & 0.2580$\dagger$* & \textbf{0.5499} & 0.9768 & 0.6501$\dagger$* & & 0.1892$\dagger$* & \textbf{0.6332} & 0.8955 & 0.5210$\dagger$* \\ %
    \rule[-2ex]{0pt}{0pt}g & RPRS w/freq  & R TSDAE & \textbf{0.2663}$\dagger$* & \textbf{0.5499} & \textbf{0.9774} & \textbf{0.6676}$\dagger$* & & \textbf{0.2015}$\dagger$* & \textbf{0.6332} & \textbf{0.8980} & \textbf{0.5360}$\dagger$* \\ \hline
    \end{tabular}
}
\end{table}
\section{Further analysis and discussion}\label{sec:discussion}
In this section, we further analyze our results, starting with an ablation study on the components of proposed methods, followed by the effect of document length, document coverage, the parameter space, and a more detailed comparison to the SDR baseline.
\begin{table}[t]
\caption{Ablation study results on the COLIEE 2021 legal case retrieval task.} %
\begin{tabular}{llccc} \label{tab:ablation}
                                 & & P           & R               & F1              \\ \hline
\rule[-2ex]{0pt}{0pt}\rule{0pt}{3ex}&Full method (\hyperref[sec:freq]{RPRS w/freq}) & \textbf{0.1960} & \textbf{0.2891} & \textbf{0.2336} \\ \hline
\rule{0pt}{3ex}a &  No QP (Eq.\ref{eq:qp}) %
& 0.1670 & 0.2421 & 0.1976 \\
b &  No DP (Eq.\ref{eq:dp}) %
& 0.1410 & 0.2180 & 0.1712 \\
\rule[-2ex]{0pt}{0pt}c &  No b and k1 params %
& 0.1860 & 0.2691 & 0.2199 \\ \hline
\rule[-2ex]{0pt}{0pt}\rule{0pt}{3ex}& RPRS without freq (Eq.\ref{eq:prs}) & 0.1890 & 0.2799 & 0.2256 \\
\rule[-2ex]{0pt}{0pt}d &  No min function in \hyperref[sec:proposed_method_pb]{PRS} & 0.1798 & 0.2644 & 0.2140 \\\hline
\end{tabular}
\end{table}
\subsection{Ablation Study on RPRS w/freq components} \label{sec:ablation}
The contribution of each module in RPRS w/freq is assessed by an ablation study in Table \ref{tab:ablation}. The role of Query Proportion (QP), Document Proportion (DP), and the parameters ($b$ and $k1$) has been evaluated in rows $a$ (No QP), $b$ (No DP), $c$ (no b and k1). 
For the sake of clarity, we re-write the modified eqution according to each row. For row $a$ (No QP), we remove the equation \ref{eq:qp} from \emph{RPRS w/freq} formula:
\begin{equation*}
    RPRS \, w/freq = DP
\end{equation*}
Similarly for row $b$ (No DP):
\begin{equation*}
    RPRS \, w/freq = QP
\end{equation*}
To do the ablation study of row $c$ (no b and k1), we modify equation \ref{eq:fqrn}, $Fq_{R_n}$, and equation \ref{eq:fdrn}, $Fd_{R_n}$,  as follow:
\begin{equation*}
    Fq_{R_n}(S_q,S_d,S_{TK_q}) = \frac{\sum_{q_{s}}^{S_q} |S_d \cap r_n(q_{s},S_{TK_q})|}{\sum_{q_{s}}^{S_q} |S_d \cap r_n(q_{s},S_{TK_q})|}
\end{equation*}
\par
\begin{equation*}
    Fd_{R_n}(S_d,S_q,S_{TK_q}) = \frac{\sum_{d_{s}}^{S_d} \sum_{r_n}^{R_n} | \{d_{s}\} \cap r_n|}{\sum_{d_{s}}^{S_d} \sum_{r_n}^{R_n} | \{d_{s}\} \cap r_n|}
\end{equation*}
The first and second highest drop in quality of ranking is caused by removing $DP$ and $QP$, which shows the crucial role of both $QP$ and $DP$ in our re-ranker. This aligns with the assumption for designing our re-ranker: a candidate document and query are likely relevant if a large proportion of their textual content is similar to each other. We further analyse the effect of the min function in $RPRS$ for the equations of $q_{R_n}$ (Eq \ref{eq:drn}) and $d_{R_n}$ (Eq \ref{eq:qrn}), in row $d$. To do so, we modify equation \ref{eq:qrn}, $q_{R_n}$, and \ref{eq:drn}, $q_{R_n}$, as follows:
\begin{equation*}
    q_{R_n}(S_q,S_d,S_{TK_q}) = 
    \sum_{q_{s}}^{S_q} |S_d \cap r_n(q_{s},S_{TK_q})|
\end{equation*}
\begin{equation*}
    d_{R_n}(S_d,S_q,S_{TK_q}) = \sum_{d_{s}}^{S_d} \sum_{r_n}^{R_n} | \{d_{s}\} \cap r_n|
\end{equation*}
In general, Table \ref{tab:ablation} shows that the full method with all components including QP, DP, and $b$ and $k1$ parameters obtains highest effectiveness and supports the necessity of each component. Morover, it ranks the importance and impact of each component in the effectiveness of the method. The importance of the $min$ function in $RPRS$ supports the intuition that if the repetition of sentences of a candidate document in $R_n$ is not controlled, the quality of the RPRS drops. This supports the $min$ function for \emph{RPRS} and motivate proposing \emph{RPRS w/freq} to take into account the repetitions in an intelligent way by adding $k1$ and $b$.
\begin{figure}[tbp]
\scriptsize
\centering
\scalebox{.6}{\includegraphics[]{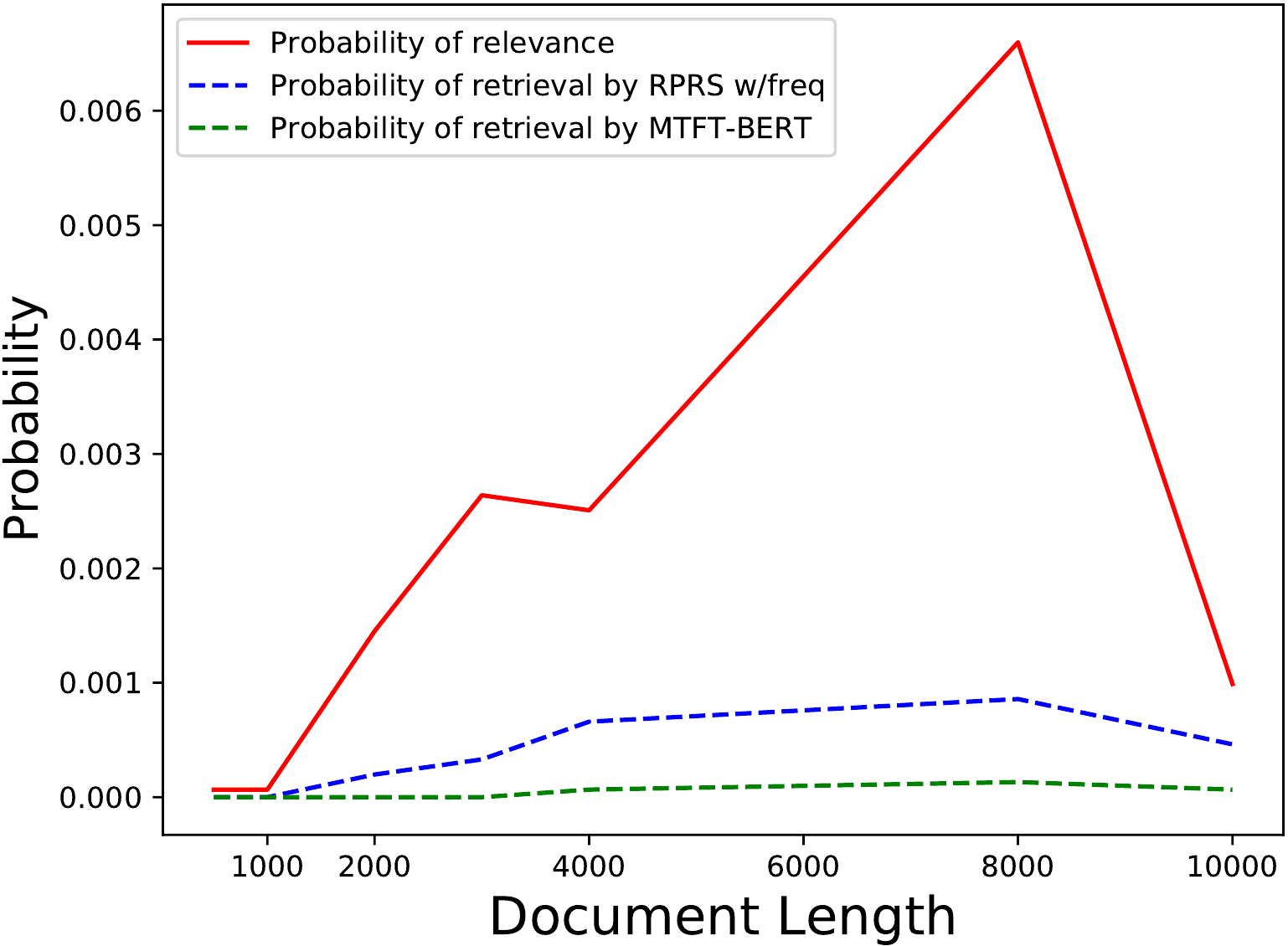}}%
\caption{The probability of relevance and probability of retrieval on COLIEE 2021. In comparison to the probability of relevance, longer documents have a disproportionately smaller probability of being retrieved by MTFT-BERT while \emph{RPRS w/freq} is not biased against retrieving long documents.}%
\label{fig:analysis_length_effect}
\end{figure}
\subsection{Effect of document length in comparison to MTFT-BERT}
We plot Figure \ref{fig:analysis_length_effect} in order to analyze the effect of document length on the effectiveness of the proposed method, and compare it with the MTFT-BERT model on COLIEE21. This comparison is justified due to the fact that while in MTFT-BERT the document length is bounded by the maximum input length of BERT, it is the current state-of-the-art model on COLIEE'21. Therefore, we select the most effective available model on this dataset for comparison in our analysis. In Figure \ref{fig:analysis_length_effect},
probability of relevance $P(relevant|length(doc))$ indicates the chance of having a relevant document with a specific length among all documents, and probability of retrieval $P(ret|length(doc))$ shows the chance of retrieving a relevant document with a specific length among all retrieved documents, within top-$k$ ranks. To compute the probability of retrieval, we set $k=5$ because the average number of relevant documents per query in the collection is five.
We analyze the \textit{probability of retrieval} and \textit{probability of relevance} from relatively shorter documents -- with $1$,$000$ words -- towards longer documents with $10$,$000$ words. An ideal ranker does not lose effectiveness, probability of retrieval in Figure \ref{fig:analysis_length_effect}, by the increment in the length of documents. As shown in Figure \ref{fig:analysis_length_effect}, the \emph{RPRS w/freq} model not only does not loose effectiveness with increasing the document length, but also retrieves the longer relevant documents easier and gets closer to the probability of relevance $P(relevant|length(doc))$ for them. This could be due to the fact that longer documents provide more information, and the \emph{RPRS w/freq} model can take use that information effectively for ranking. This is while MTFT-BERT does not improve by increasing document length and receive similar effectiveness even with the more information that exist in a longer document.
This indicates that one of the reasons for gaining a higher effectiveness of F1 score, which is the official metric of COLIEE dataset, by the proposed method compared to the other models could be because of the fact that \emph{RPRS w/freq} could capture the full information in the lengthy documents and retrieves them more effective than other models such as MTFT-BERT. 
\subsection{Effect of covering the full length of queries and documents}
\begin{figure}[t]
\scriptsize
\centering
\scalebox{.65}{\includegraphics[]{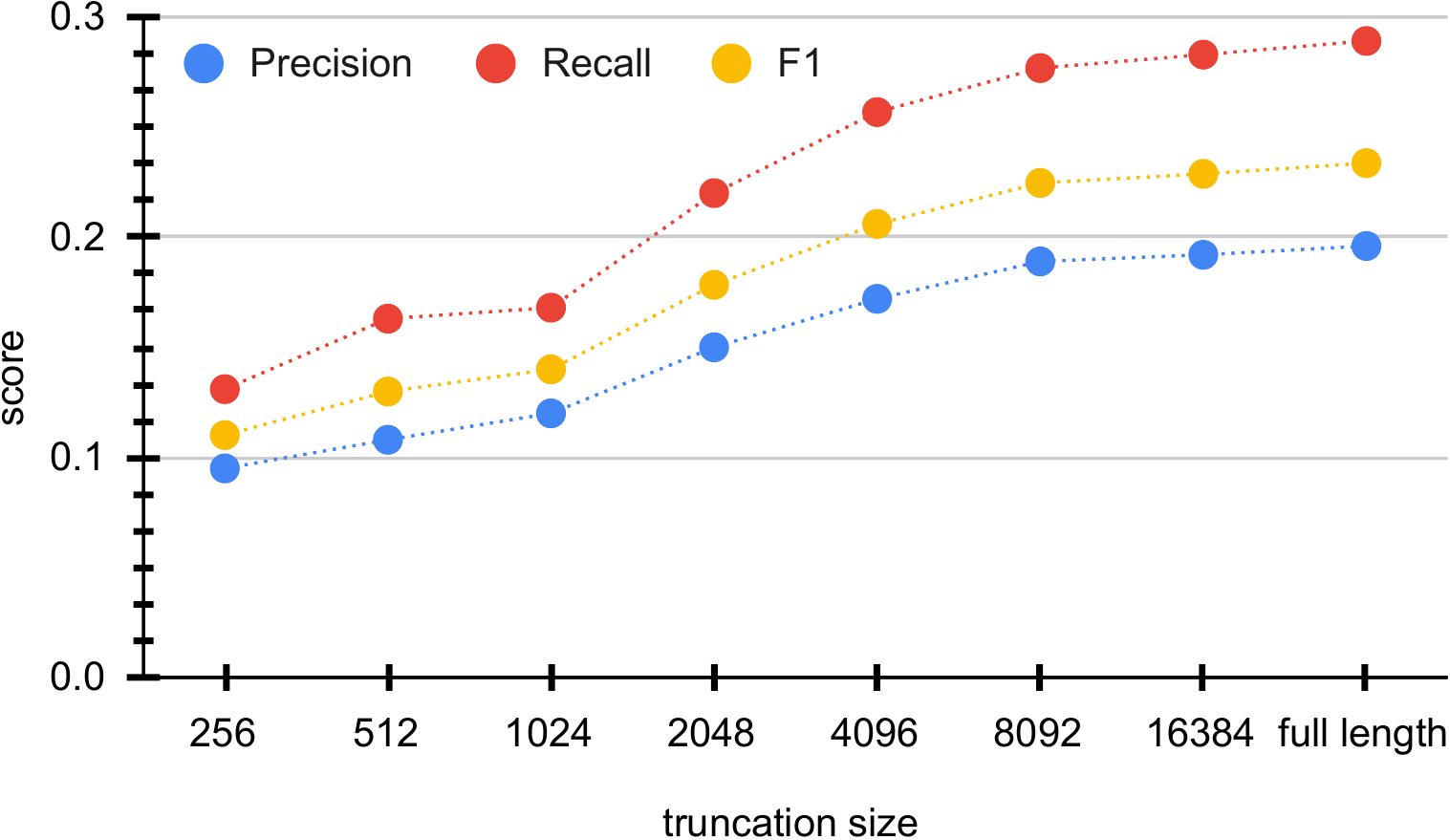}}
\caption{Effectiveness of RPRS w/freq over varying truncation length for queries and documents on COLIEE 2021 where truncation size is the maximum length input.}%
\label{fig:analysis_truncation_effect}
\end{figure}

We analyze if taking into account the full length of queries and documents is an advantage for the proposed method in Figure \ref{fig:analysis_truncation_effect}. We argue that an ideal semantic re-ranker should achieve higher effectiveness by receiving the full length of queries and documents rather than truncated text as the input. Therefore, we analyze the power of our re-ranker when the length of queries and documents increases: we experiment with  different maximum lengths for the input. Given $l \in \{256,$ $512,$ $1024,$ $2048,$ $4096,$ $8092,$ $16384\}$ as the maximum input length, we select the first $l$ tokens of the query and documents as their representation. 
Moreover, our analysis gives an in-depth insight about the robustness of the proposed method and assess whether it collapse when presented with long queries and documents or not. As Figure \ref{fig:analysis_truncation_effect} shows, our re-ranker takes advantage of seeing the whole content of query and documents in terms of effectiveness as it has the highest effectiveness when we feed it with the full length and the lowest effectiveness with only $256$ tokens as the input. The figure also shows the largest leap is between 1024 and 4096 tokens, which is approaching the average document length in COLIEE ($5226$). This confirms that $RPRS w/freq$ takes advantage of the full document length in estimating relevance.
\subsection{Parameter sensitivity for $RPRS \, w/freq$}
In addition to the tuned parameters that were found and explained in Section \ref{sec:parameter_tuning}, we analyze the sensitivity of the proposed parameters in other ranges in the following. Figure \ref{fig:sensitivity} shows the effects of changes in parameters $n, b$ and $k1$ on the overall performance of our method on the COLIEE 2021 dataset. In each diagram, the value of each point represents the F1 score according to different values of the parameters. The following observations can be made from Figure \ref{fig:sensitivity}:
\begin{figure}[tbh]
    \centering %
\begin{subfigure}{0.33\textwidth}
  \includegraphics[width=\linewidth]{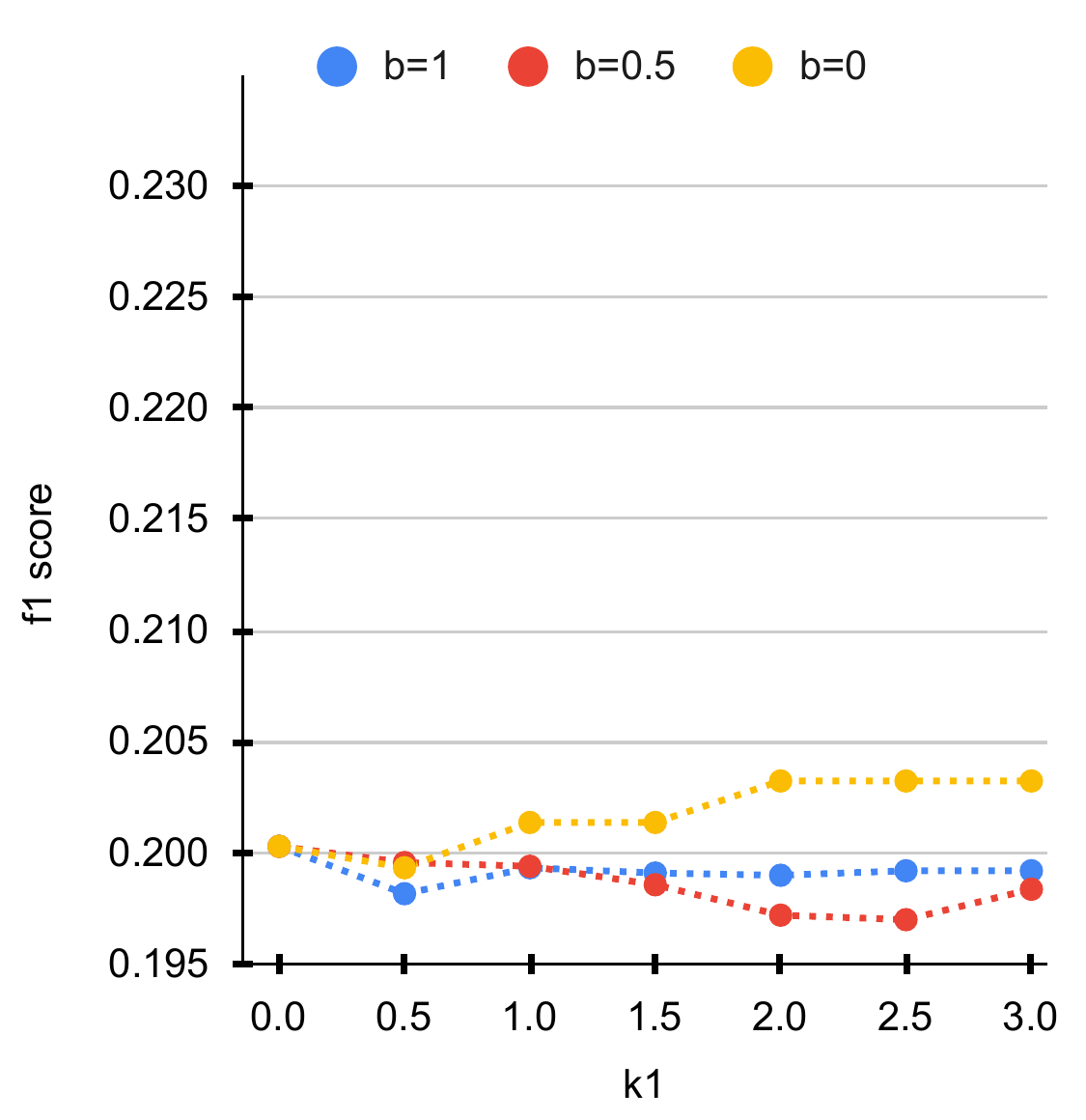}
  \caption{n=1}%
  \label{fig:sensitivity_n1}
\end{subfigure}\hfil %
\begin{subfigure}{0.33\textwidth}
  \includegraphics[width=\linewidth]{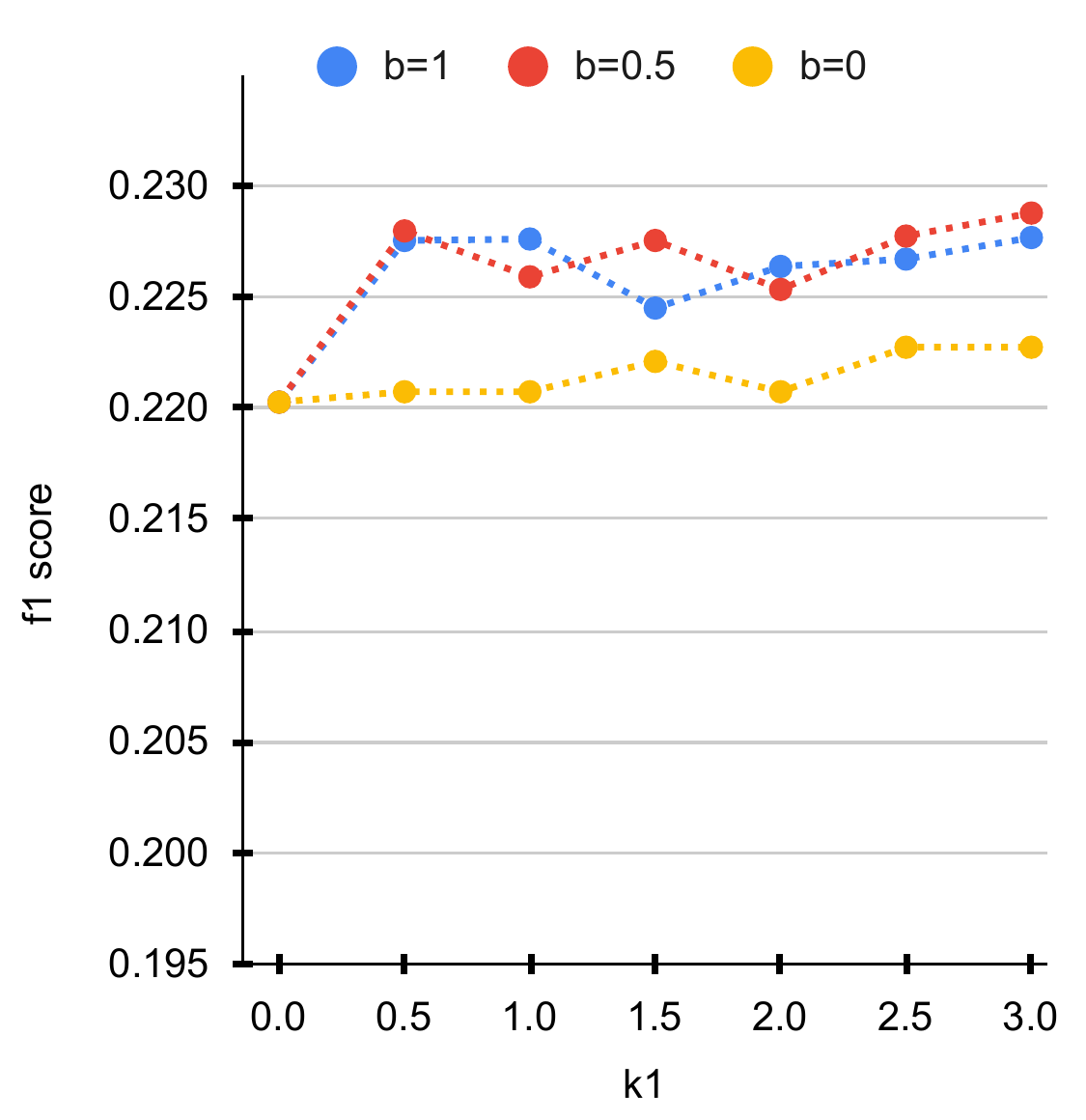}
  \caption{n=5}%
  \label{fig:sensitivity_n5}
\end{subfigure}\hfil %
\begin{subfigure}{0.33\textwidth}
  \includegraphics[width=\linewidth]{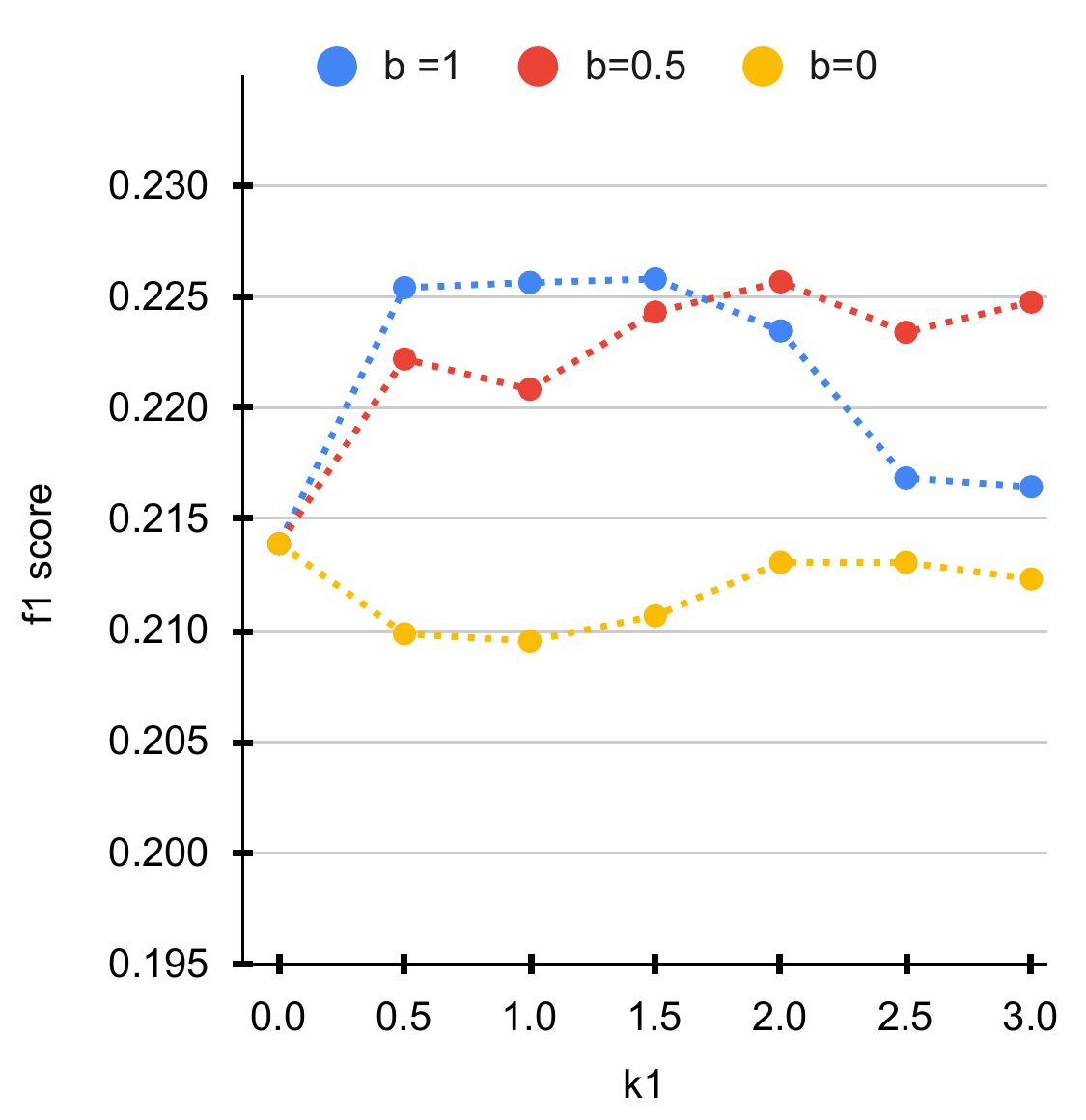}
  \caption{n=10}%
  \label{fig:sensitivity_n10}
\end{subfigure}
\caption{Sensitivity analysis of our proposed method, $RPRS w/freq$, to changes in parameters n, b and k1 on the COLIEE 2021 dataset.}
\label{fig:sensitivity}
\end{figure}
\begin{itemize} 
    \item A comparison between different values of $n$ -- independent from $b$ and $k1$ -- indicates that the proposed method has the lowest effectiveness by reducing the value of $n$ to $1$. The effectiveness of the model increases as $n$ is set to $5$, and as we set $n=10$, we see that model effectiveness declines while still being superior to $n$ set to $1$. We argue that this demonstrates the existence of a trade off in the value of $n$, which determines the number of the most similar sentences per query sentence in set $r_n$. If $n$ is set to $1$, the selection of \textit{most similar} sentences per query sentence would be very strict, and the effectiveness would decrease. This could be because that there may be other sentences from documents that are similar enough to qualify as a \textit{most similar} sentence to a query sentence  — in addition to the top-$1$ most similar sentence — but with $n=1$, they will not be taken into account in the computation of relevance score. On the other hand, if $n$ is set to $10$, some sentences that are not similar enough to qualify as a \textit{most similar} sentence to a query sentence may appear in $r_n$, and as the result the model obtain less effectiveness compared to $n=5$ because it would be less strict in creation of $r_n$s for query sentences.
    \item If the $k1$ parameter is set to $0$ for any value of the $b$ parameter, the $f1$ score depends only on the value of the $n$ parameter. This is due to the fact that $k1$ multiplies to the $b$ parameter, and if $k1$ is set to zero, the relevance score is independent of changes in the $b$ parameter's value. Taking this into consideration, we could reduce the search space for parameter optimization by eliminating the necessity of grid search for $b$ values when $k1$ is set to zero.
    \item By increasing the value of the $n$ parameter the proposed method becomes more sensitive to the document length normalization parameter $b$, and frequency saturation parameter $k1$.
    There are larger disparities in the $f1$ scores when the $k1$ or $b$ parameter are changed for $n=10$ than for $n=5$, and for $n=5$ than for $n=1$. This indicates where the method is less strict in qualifying a sentence as a \textit{most similar} sentence to a query sentence, the impact of term saturation parameter ($k1$) and document length normalization ($b$) are higher.
    \item The proposed method performs better in terms of effectiveness without doing length normalization ($b = 0$) when $n$ is set to $1$, and, as the result, the selection of the most similar sentences per query sentence is very strict, and it performs better with length normalization ($b=0.5$ and $b=1.0$) when $n$ is less strict ($n=5$ and $n=10$). This could probably shows that the negative effects on effectiveness by less strictly selecting the most similar sentences per query sentence --  by increasing the value of $n$ parameter -- could might be controlled with $b$ parameter.
\end{itemize}
In general, the parameter sensitivity analysis reveals that while optimizing RPRS parameters increases effectiveness, it can still result in increased effectiveness even without optimization by initializing each of the three parameters with the value from middle of their range, e.g., $n=5$, $k1=1.5$, $b=0.5$ that obtains $22.75$ f1 score which is still better than current state-of-the-art on the dataset (MTFT-BERT\cite{AbolghasemiECIR22}) while not being the ideal result by our \emph{RPRS w/freq} re-ranker.
\subsection{Further analysis of RPRS compared to SDR}
We compared RPRS to SDR because it has some similarities: it also uses sentence-level relevance scoring based on sentence embeddings and it uses a bi-encoder architecture, like RPRS. There are fundamentally significance differences, however: SDR creates a sentence-level similarity matrix $M$ for each pair of query and document paragraphs and based on $M$ matrices, creates a paragraph similarity matrix $P$ for a candidate document. Each cell of $P$ contains the similarity between a pair of paragraphs from query and a candidate document. The matrix $P$ is then normalized to ($NRM(P)$) and the total score $S$ is computed based on $NRM(P)$.
The denominator in the formula for computing the matrix $P$ is the number of sentences in a query paragraph and the denominator for the total score $S$ is the number of paragraphs in the query. An important consequence is that the length of the candidate documents is not taken into account, while RPRS explicitly does so. Moreover, if one paragraph of a candidate document is the most similar paragraph to all paragraphs of the query, SDR does not penalize this repetition, while we take that into account with the $k1$ parameter in \emph{RPRS w/freq}, which controls frequency saturation. 
In addition, it is noteworthy to mention that SDR uses the cosine similarity directly in their approach while we only consider that score for ranking sentences in $R_n$.
\par
\begin{table}[]
\centering
\caption{Pearson Product-Moment Correlation coefficients between document length and relevance score.}
\label{table:pearson}
\begin{tabular}{ll!{\color[rgb]{0.753,0.753,0.753}\vrule}ccccc} 
\hline
\rule[-2ex]{0pt}{0pt}\rule{0pt}{3ex}\, &             & \multicolumn{5}{c}{Pearson Product-Moment Correlation coefficient}  \\
\rule[-2ex]{0pt}{0pt}\, & Model       & COLIEE 2021 & Caselaw & CLEF-IP & Video Games & Wines datasets      \\ 
\hline
 \rule[-2ex]{0pt}{0pt}\rule{0pt}{3ex}\,& SDR         & $0.5768$      & $0.5901$  & $0.612$   & $0.542$       & $0.5366$              \\
 \rule[-2ex]{0pt}{0pt}\,& RPRS w/freq & $- 0.0565$  & $- 0.0401$ & $0.0105$  & $- 0.0322$     & $0.0109$              \\
\hline
\end{tabular}
\end{table}
\begin{figure}[t]
\scriptsize
\centering
\scalebox{.5}{\includegraphics[]{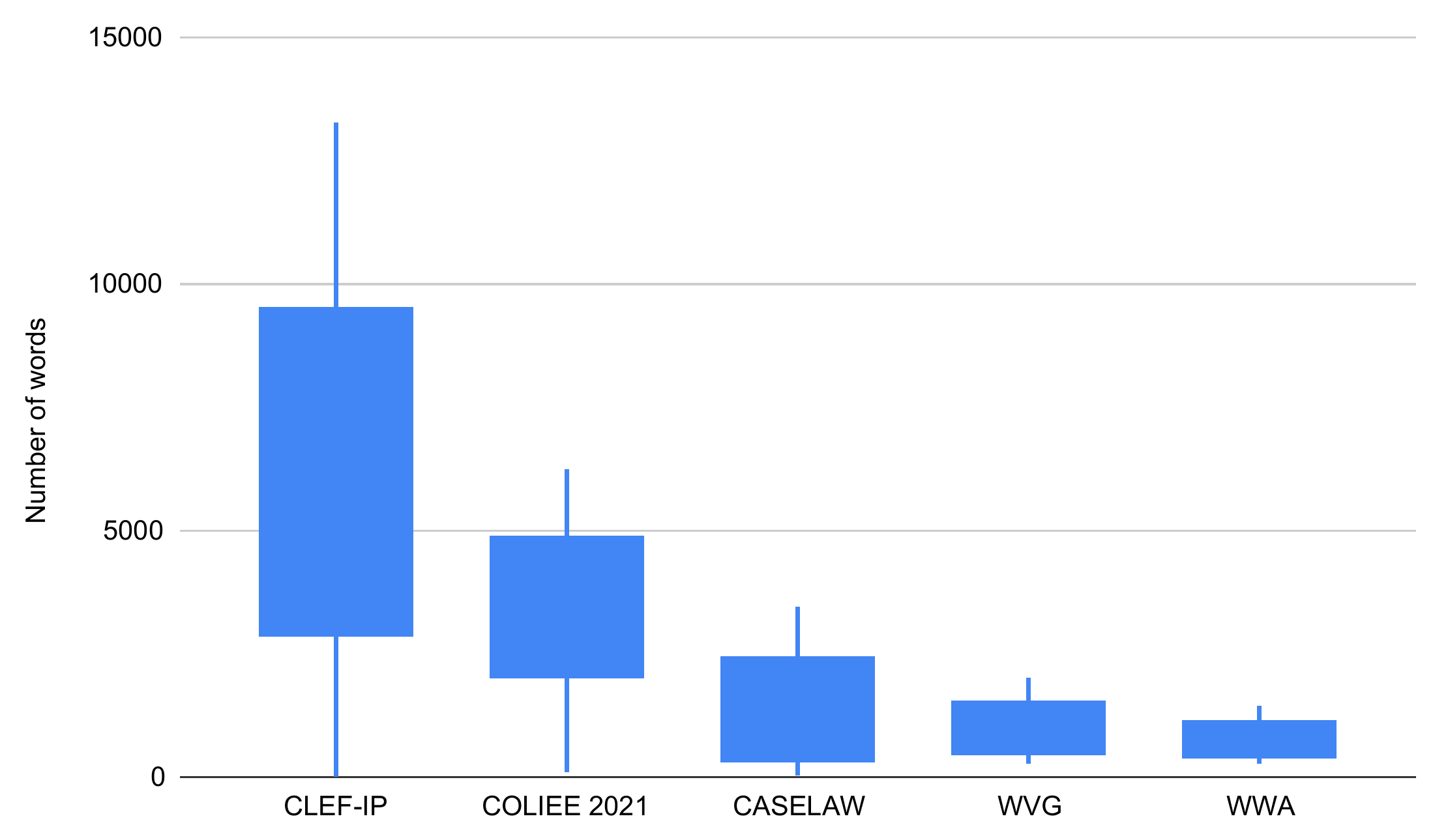}}
\caption{Distribution of word counts per each document for all of the five datasets: Patent retrieval (CLEF-IP), Case law retrieval (COLIEE 2021 and CASELAW), and Wikipedia (WWG and WWA).}%
\label{fig:distribution_length}
\end{figure}
Losada et al. \cite{losada2008revisiting} indicate that an ideal retrieval model should not be tuned to favour longer documents and the relevance scores it produces should not be correlated with document length. They argue that the previous empirical evidence supporting the \textit{scope hypothesis}\footnote{The scope hypothesis in Information Retrieval states that a relationship exists between document length and relevance \cite{robertson1994some}.} is over-exaggerated and inaccurate due to the incompleteness of modern collections. 
This is even more important in QBD retrieval as the length of relevant documents could vary from short  to very long. Therefore, the retrieval model should not be biased to the document length.
Consequently, in order to study the sensitivity of our proposed method to document length, and compare it with SDR, we analyze the correlation between candidate document length and the relevance score produced by \emph{RPRS w/freq} and SDR in Table \ref{table:pearson}. We find that the Pearson Product-Moment Correlation coefficients for \emph{RPRS w/freq} on all five datasets are close to zero while for \emph{SDR} these correlations are much higher, above $0.5$, for all datasets. This indicates the strong correlation of SDR relevance score with the document length and on the other hand the robustness of \emph{RPRS w/freq} with respect to document length. In other words, \emph{RPRS w/freq} is effective while it is not biased to the length of the document.
\par
In addition to the correlation analysis, we further study the statistics over word counts of documents in all datasets to find out if there is a specific statistical characteristic in the Video games and Wines datasets that makes SDR stronger on those datasets compared to the three datasets in the legal and patent domain. Figure \ref{fig:distribution_length} shows the document length distribution over the datasets that indicates higher variance among Patent and Legal datasets compared to Wikipedia datasets (WVG and WWA). The boxes bound the $25$th to $75$th percentiles, top whisker cover data within $1.5 \times$ the inter-quartile range (IQR), and outliers are removed. Moreover, the standard deviation of documents length for the Video Games and Wines datasets are 675 abd  $721$ respectively. On the other hand, for COLIEE 2021, Caselaw, and CLEF-IP, the standard deviation  of documents length are $6933$, $4937$, and $11402$. Thus, the documents in SDR's Wikipedia datasets have less variance compared to the three legal and patent datasets and are shorten than COLIEE and Patent datasets. As a result, we conclude that the lower effectiveness of SDR on the legal and patent datasets is likely caused by SDR not being robust against the length of documents and suffers from the high standard deviation, and longer documents in the legal and patent datasets in comparison to its own datasets. Our experimental results in Section~\ref{sec:generalizability} indicate that our method \emph{RPRS w/freq} also outperforms SDR on their datasets, so the robustness of \emph{RPRS w/freq} is not only beneficial for extremely lengthy documents but also in other domains.

\subsection{Effectiveness on different versions of the COLIEE dataset}
In order to analyze the effectiveness of the proposed methods in more depth, we investigate the effectiveness of the proposed methods compared to the previous state-of-the-art methods on different versions of the COLIEE datasets. We analyze this on COLIEE 2020 and COLIEE 2022. Table \ref{tab:res_coliee2020} and \ref{tab:res_coliee2022} shows that our re-ranker outperforms the state-of-the-art model on both versions of COLIEE for all evaluation metrics with or without taking into account frequency (lines $d$ and $e$). Row $d$ of Tables \ref{tab:res_patent} show that even without taking into account frequency, \emph{RPRS} achieves higher effectiveness than the previous state-of-the-art method ($b$). Paraformer \citet{nguyen2022attentive} reports F2 in their paper while F1 is the official metric for COLIEE dataset. Therefore, we report F1 instead of F2 in Table \ref{tab:res_coliee2020}.\footnote{It is important to note that \citet{nguyen2022attentive} focus on COLIEE 2020 and report all the baselines on COLIEE 2020 while mistakenly mention COLIEE 2021 as the used dataset in their experiments.}
\begin{table}[]
\centering
\caption{Results for COLIEE’2020. $\dagger$ and * indicate a statistically significant improvement over the previous state-of-the-art (JNLP team \cite{nguyen2020jnlp}, row b) and BM25 + KLI (line a) respectively, according to a paired-test (p<0.005) with Bonferroni correction for multiple testing. The winning team in the COLIEE 2020 competition for legal case retrieval is JNLP \cite{nguyen2020jnlp}. The initial ranker is BM25\textsubscript{optimized}+KLI. Legal SBERT-TSDAE is used to embed sentences for RPRS, and \emph{RPRS w/freq}.}
\label{tab:res_coliee2020}
\arrayrulecolor[rgb]{0.753,0.753,0.753}
\begin{tabular}{lll|c|c|c} 
\arrayrulecolor{black}\hline
 \rule[-2ex]{0pt}{0pt}\rule{0pt}{3ex}& & Model & Precision & Recall & F1 \\ 
\arrayrulecolor[rgb]{0.502,0.502,0.502}\hline
 \rule[-2ex]{0pt}{0pt}\rule{0pt}{3ex}& \multicolumn{2}{l!{\color{silver}\vrule}}{\textbf{Probabilistic lexical matching baselines}} & \multicolumn{1}{l!{\color{silver}\vrule}}{} & \multicolumn{1}{l!{\color{silver}\vrule}}{} & \multicolumn{1}{l}{} \\
 & a & BM25 + KLI & 0.6700 & 0.6117 & 0.6395 \\ \hline
  \rule[-2ex]{0pt}{0pt}\rule{0pt}{3ex}& \multicolumn{2}{l!{\color{silver}\vrule}}{\textbf{Baselines}} & \multicolumn{1}{l!{\color{silver}\vrule}}{} & \multicolumn{1}{l!{\color{silver}\vrule}}{} & \multicolumn{1}{l}{} \\
 \rule[-2ex]{0pt}{0pt}& b & \multicolumn{1}{l!{\color{silver}\vrule}}{JNLP team \cite{nguyen2020jnlp} (Previous state-of-the-art)} & 0.8025 & 0.7227 & 0.7605 \\ 
  \rule[-2ex]{0pt}{0pt}& c & \multicolumn{1}{l!{\color{silver}\vrule}}{Paraformer \cite{nguyen2022attentive} } & 0.7346 & 0.7407 & 0.7376 \\ 
\hline
\arrayrulecolor[rgb]{0.502,0.502,0.502}\hline
 \rule[-2ex]{0pt}{0pt}\rule{0pt}{3ex}& \multicolumn{2}{l!{\color{silver}\vrule}}{\textbf{\textbf{\textbf{\textbf{Proposed methods}}}}} & \multicolumn{1}{l!{\color{silver}\vrule}}{} & \multicolumn{1}{l!{\color{silver}\vrule}}{} & \multicolumn{1}{l}{} \\
 & d & RPRS without frequency & 0.7840 & 0.7550 & 0.7692 \\
 \rule[-2ex]{0pt}{0pt}& e & RPRS w/freq & 0.7980 & \textbf{0.7690$\dagger$*} & \textbf{\textbf{0.7832$\dagger$*}} \\
\arrayrulecolor{black}\hline
\end{tabular}
\end{table}
\begin{table}[]
\centering
\caption{Results for COLIEE’2022. $\dagger$ and * indicate a statistically significant improvement over the previous state-of-the-art (UA team \cite{rabelo2022semantic}, row b) and BM25 + KLI (line a) respectively, according to a paired-test (p<0.005) with Bonferroni correction for multiple testing. The winning team in the COLIEE 2020 competition for legal case retrieval is UA \cite{rabelo2022semantic}. The initial ranker is BM25\textsubscript{optimized}+KLI. Legal SBERT-TSDAE is used to embed sentences for RPRS, and \emph{RPRS w/freq}.}
\label{tab:res_coliee2022}
\arrayrulecolor[rgb]{0.753,0.753,0.753}
\begin{tabular}{lll|c|c|c} 
\arrayrulecolor{black}\hline
 \rule[-2ex]{0pt}{0pt}\rule{0pt}{3ex}& & Model & Precision & Recall & F1 \\ 
\arrayrulecolor[rgb]{0.502,0.502,0.502}\hline
 \rule[-2ex]{0pt}{0pt}\rule{0pt}{3ex}& \multicolumn{2}{l!{\color{silver}\vrule}}{\textbf{Probabilistic lexical matching baselines}} & \multicolumn{1}{l!{\color{silver}\vrule}}{} & \multicolumn{1}{l!{\color{silver}\vrule}}{} & \multicolumn{1}{l}{} \\
 & a & BM25 + KLI & 0.3000 & 0.2850 & 0.2923 \\ \hline
  \rule[-2ex]{0pt}{0pt}\rule{0pt}{3ex}& \multicolumn{2}{l!{\color{silver}\vrule}}{\textbf{Baselines}} & \multicolumn{1}{l!{\color{silver}\vrule}}{} & \multicolumn{1}{l!{\color{silver}\vrule}}{} & \multicolumn{1}{l}{} \\
 \rule[-2ex]{0pt}{0pt}& b & \multicolumn{1}{l!{\color{silver}\vrule}}{UA team \cite{rabelo2022semantic} (Previous state-of-the-art)} & 0.4111 & 0.3389 & 0.3715 \\ 
  \rule[-2ex]{0pt}{0pt}& c & \multicolumn{1}{l!{\color{silver}\vrule}}{Siat team \cite{wen2022siat} } & 0.3005 & 0.4782 & 0.3691 \\ 
\hline
\arrayrulecolor[rgb]{0.502,0.502,0.502}\hline
 \rule[-2ex]{0pt}{0pt}\rule{0pt}{3ex}& \multicolumn{2}{l!{\color{silver}\vrule}}{\textbf{\textbf{\textbf{\textbf{Proposed methods}}}}} & \multicolumn{1}{l!{\color{silver}\vrule}}{} & \multicolumn{1}{l!{\color{silver}\vrule}}{} & \multicolumn{1}{l}{} \\
 & d & RPRS without frequency & 0.4244 & 0.3607 & 0.3900 \\
 \rule[-2ex]{0pt}{0pt}& e & RPRS w/freq & \textbf{\textbf{0.4361$\dagger$*}} & 0.3904 & \textbf{\textbf{0.4120$\dagger$*}} \\
\arrayrulecolor{black}\hline
\end{tabular}
\end{table}

\section{Conclusion and Future work}\label{sec:conclusion}
In this paper, we proposed methods for effectively exploiting sentence-level representations produced by the highly efficient bi-encoder architecture for Query-by-Document (QBD) re-ranking. We proposed a novel model using SBERT representations, with a frequency-based extension inspired by BM25's `term saturation' mechanism and the incorporation of document length normalization into the relevance score computation. Our experiments on five datasets show that our model \emph{RPRS w/freq} takes advantage of the long queries and documents that are common in QBD retrieval. While our \emph{RPRS w/freq} model is unsupervised with only three tunable parameters, it is more effective than state-of-the-art supervised neural and lexical models. In addition, it is highly efficient for retrieval tasks with long documents and long queries because the operation in order to compute relevance score by \emph{RPRS w/freq} are based on: (1) a bi-encoder, SBERT, which is about $46,800$ times faster than the common cross-encoder BERT architecture; (2) the pre-processing, embedding, and indexing of document sentences could be done before the query time; (3) the only calculation based on the embeddings are cosine similarity and sorting that are simple and efficient operations.
\par
We show the effectiveness of \textit{RPRS w/freq} on five datasets with low-resource training data, which indicates its suitability for QBD retrieval tasks in which the training data is very limited compared to general web search due to the high cost of dataset creation for these tasks. Therefore, we attain high efficiency and effectiveness while being optimized on low-recourse training data with \emph{RPRS w/freq}. 

While we outperform the state-of-the-art models on each dataset with our proposed method, the effectiveness results show the difficult nature of the tasks. One reason for that is the low effectiveness of the first stage retrieval model limits the re-ranker performance. As one direction of improvement for future work, we aim to focus on first-stage retrieval by designing a modification to our proposed method which is suitable for first-stage retrieval tasks. The parameter sensitivity analysis reveals that while optimizing RPRS parameters increases effectiveness, it can still result in increased effectiveness even without optimization by initializing each of the three parameters with a value from the middle of their range.
\par
For further future improvement on our model, we might be inspired by two other variants of BM25 to make the proposed method parameter-free with dynamically computed parameters. For the frequency saturation parameter, $k1$, a variant of BM25 called `BM25 adaptive' \cite{lv2011adaptive} which dynamically computes the $k1$ parameter could be taken into account as inspiration. Additionally, for the document length normalization parameter, $b$, Lipani et al. \cite{lipani2015verboseness} propose a length normalization method that removes the need for a $b$ parameter in BM25 which could inspire us to dynamically compute the $b$ parameter of \emph{RPRS w/freq} in future. Considering the intuition of both approaches, working on dynamically initializing the $n$ parameter could make our method completely parameter-free which is an interesting direction for future work. We argue because a sensitivity analysis reveals a pattern that indicates robust and greater effectiveness relative to the baseline for the proposed method with different parameter values, finding a nearly optimal parameter value dynamically might likely be possible.
\par
We believe the proposed method is suitable for other tasks that deal with long documents. In addition, we suggest analyzing the effectiveness of our proposed method on other IR tasks in which we do not have extremely long documents, but the query and candidate document do consist of multiple sentences \cite{cohan2020specter,AbolghasemiIctir2022}. We hope that our work will spark more research interest in retrieval for extremely long queries and open up possibilities for highly efficient long-document retrieval. 
\bibliographystyle{ACM-Reference-Format}
\bibliography{main}
\end{document}